\documentclass[12pt]{article}

\pdfoutput=1 
\usepackage{mathrsfs}
\usepackage{graphicx}
\usepackage{xcolor}
\usepackage{cancel,tabularx,moreverb,fancybox}
\usepackage{amsmath,amssymb,bm,braket,txfonts,accents}
\usepackage{float}
\usepackage[top=30truemm,bottom=30truemm,left=25truemm,right=25truemm]{geometry}
\usepackage{latexsym}
\usepackage{here}
\usepackage{cite}
\usepackage{hyperref}
\newcommand{\ex}[1]{\mathrm{e}^{#1}}
\newcommand{\fr}{\frac}
\newcommand{\pa}[1]{\left(#1 \right)}
\newcommand{\ca}[1]{\mathcal{#1}}
\newcommand{\bb}[1]{\mathbb{#1}}

\def\del{{\partial}}
  \makeatletter
    
    \@addtoreset{equation}{section}
  \makeatother
\begin{document}

\begin{titlepage}
\thispagestyle{empty}

\begin{flushright}
KYUSHU-HET-364
\\
RIKEN-iTHEMS-Report-26
\\

\end{flushright}

\bigskip

\begin{center}
\noindent{{\large \textbf{
Complex Conformal Manifolds
}}}\\
\vspace{2cm}
Yuma Furuta $^1$, Wataru Harada ${}^2$, Yuya Kusuki ${}^{1,2,3}$, Yin Tang ${}^1$
\vspace{1cm}

${}^{1}${\small \sl 
Institute for Advanced Study, \\
Kyushu University, Fukuoka 819-0395, Japan
}

${}^{2}${\small \sl 
Department of Physics, \\
Kyushu University, Fukuoka 819-0395, Japan
}

${}^{3}${\small \sl RIKEN Interdisciplinary Theoretical and Mathematical Sciences (iTHEMS), \\Wako, Saitama 351-0198, Japan}

\vskip 2em
\end{center}

\begin{abstract}
Complex conformal field theories (CFTs) have recently emerged as essential frameworks for understanding non-Hermitian criticality, weakly first-order phase transitions, and walking renormalization group flows, while their general structures remain largely unknown. In this work, we propose a systematic construction of complex CFTs by analytically continuing exactly marginal couplings into the complex plane. This procedure applies uniformly to bulk, boundary, and defect deformations, preserving conformal symmetry while generically complexifying operator spectra and other universal data.
Using the compact free boson as a solvable laboratory, we uncover the global structure of the complexified Gaussian conformal manifold. More generally, we demonstrate that genuinely complex rational CFTs do not exist: rational points remain confined to the real regime, providing a sharp distinction between real and complex theories. 
In the defect case, we investigate the one-parameter family of conformal defects in the Ising CFT and derive exact expressions for the defect spectrum, energy transmission coefficient, and effective central charge from analytic continuation. The theoretical predictions are precisely verified in non-Hermitian critical Ising and free fermion chains using bulk-defect correlators, entanglement entropy, and complex energy transport, providing concrete evidence for the complex defect conformal manifold.
Finally, we study complex boundary renormalization-group flows through the AdS/BCFT correspondence.
Our results establish complex conformal manifolds as a controlled bridge between solvable lattice models, complex CFTs, and holography, while providing stringent analytic benchmarks for the nonunitary conformal bootstrap.
\end{abstract}

\end{titlepage}

\tableofcontents

%%%%%%%%%%%%%%%%%%%%%%%%%%%%%%%%%%%%%%%%%%%%%%%%%%%%%%%%%%%%%%%%%%%%%%%%%%%%%%%%%%%%%%%%%%%%%%
%%%%%%%%%%%%%%%%%%%%%%%%%%%%%%%%%%%%%%%%%%%%%%%%%%%%%%%%%%%%%%%%%%%%%%%%%%%%%%%%%%%%%%%%%%%%%%
\section{Introduction \& Summary}
%%%%%%%%%%%%%%%%%%%%%%%%%%%%%%%%%%%%%%%%%%%%%%%%%%%%%%%%%%%%%%%%%%%%%%%%%%%%%%%%%%%%%%%%%%%%%%
%%%%%%%%%%%%%%%%%%%%%%%%%%%%%%%%%%%%%%%%%%%%%%%%%%%%%%%%%%%%%%%%%%%%%%%%%%%%%%%%%%%%%%%%%%%%%%

%%%%%%%%%%%%%%%%%%%%%%%%%%%%%%%%%%%%%%%%%%%%%%%%%%%%%%%%%%%%%%%%%%%%%%%%%%%%%%%%%%%%%%%%%%%%%%
\subsection{Introduction}
%%%%%%%%%%%%%%%%%%%%%%%%%%%%%%%%%%%%%%%%%%%%%%%%%%%%%%%%%%%%%%%%%%%%%%%%%%%%%%%%%%%%%%%%%%%%%%

Conformal field theory (CFT) provides one of the sharpest formulations of universality
\cite{Belavin:1984vu,DiFrancesco:1997nk,Cardy:2008jc,Rychkov:2016iqz,Poland:2018epd}.
A bulk CFT is not characterized by the microscopic Hamiltonian itself,
but by intrinsic conformal data such as operator spectra,
operator product expansion (OPE) coefficients, and central charges.
Boundaries, defects, and interfaces enrich this structure by introducing boundary
and defect sectors, with their own spectra and OPE data \cite{Cardy:1984bb,Cardy:1989ir}.
They also give rise to universal observables, such as the Affleck--Ludwig boundary entropy \cite{Affleck:1991tk},
entanglement entropy across an interface \cite{Sakai:2008tt,Eisler:2012xry,Brehm:2015lja,Brehm2015,Gutperle2015,Wen:2017smb,Rogerson:2022yim},
and reflection/transmission coefficients \cite{Quella2006,Brunner:2015vva,Meineri2019,Bachas:2020yxv,Furuta:2025ahl}.
In unitary CFTs these quantities are strongly constrained by positivity: scaling dimensions are real and bounded below, energy transmission probabilities are non-negative, and monotonicity theorems such as the $c$- and $g$-theorems rely on positivity assumptions.

However, in recent years, complex CFTs have emerged as an important framework in a wide variety of physical problems \cite{Kaplan:2009kr,Gorbenko:2018ncu}. The best-known examples arise in the two-dimensional Potts model with $Q>4$ \cite{Gorbenko:2018dtm,Ma:2018euv,Jacobsen:2024jel,Tang:2024blm} and the $O(n)$ loop model with $n>2$ \cite{Haldar:2023ukr}, where pairs of complex fixed points govern weakly first-order transitions and pseudocritical scaling. Associated walking renormalization-group (RG) flows controlled by complex fixed points have also been proposed in gauge theories near the lower edge of the conformal window \cite{Kaplan:2009kr,Benini:2019dfy} and in deconfined quantum criticality \cite{Nahum:2015jya,Wang:2017txt,Serna:2018tct,Ma:2019ysf,Nahum:2019fjw}. These developments suggest that complex CFTs are not merely mathematical curiosities, but rather provide a universal description of approximate scale invariance and long crossover phenomena in both high-energy and condensed matter systems.

More broadly, complex CFTs challenge some of the most fundamental notions in unitary CFTs, including positivity, monotonicity, and rationality. Understanding which aspects of conformal dynamics survive the loss of unitarity may reveal a deeper and more universal structure underlying quantum criticality itself.
Specifically, once the theory is analytically continued to complex
couplings, or realized microscopically by a non-Hermitian Hamiltonian, the positivity
properties need not survive. The resulting theory may still obey conformal Ward
identities and have well-defined local conformal data, but the interpretation of those
data changes. Complex scaling dimensions lead to log-periodic factors in correlation
functions; an analytically continued interface transmission coefficient need no longer admit a probabilistic interpretation; and the logarithmic coefficient in an entanglement entropy need not be real.

Despite recent developments, several fundamental questions remain open. First, most existing studies focus on complex bulk CFTs \cite{Jacobsen:2024jel,Tang:2024blm,Haldar:2023ukr,Kumar:2025zey,Yang:2026mnz,Yamamoto2026,Jacobsen:2026bvg}, while much less is known about complex boundary and defect CFTs. Moreover, the few examples of complex boundary criticality \cite{Tang:2025bju,Linden:2025nfc,Liu:2026gak} that have been studied so far are typically built upon bulk theories that already realize complex criticality, where bulk unitarity is absent from the outset. It is therefore natural to ask whether a unitary bulk CFT can admit genuinely complex conformal boundary or defect fixed points under their respective RG flows, or whether the unitarity of the bulk theory is already sufficient to enforce the unitarity of its boundary and defect sectors. A second question concerns the very origin of complex CFTs. Since complex fixed points appear to play an important role in critical phenomena, can they arise directly from familiar and exactly solvable theories? More specifically, can one start from simple CFTs and deform them into genuinely complex conformal theories while maintaining exact control over their conformal data? Such a construction would provide explicit and analytically tractable examples of complex CFTs and embed them naturally into the theory space of ordinary CFTs.

In this work, we answer the above questions within the framework of conformal manifolds \cite{Zamolodchikov1986,Cardy:1987vr,Green:2010da,Gomis:2015yaa,Karch:2018uft,Herzog:2019bom,Drukker:2022pxk,Komatsu:2025cai,Antinucci:2025uvj,Choi:2025ebk}.
Suppose that a real CFT belongs to a family generated by an exactly marginal bulk, boundary, or defect
operator, and let $\lambda$ be a local coordinate on this family.
If the beta function and conformal data are analytic near the real locus, then the identity theorem implies that the vanishing of the beta function extends to the complexified patch,
\begin{equation}
\lambda\in\mathbb{R}
\qquad\longrightarrow\qquad
\lambda\in\mathbb{C}.
\end{equation}
This produces a family of generally nonunitary but conformal theories, within the domain where the continued data remain analytic.
Although the formalism is applicable in arbitrary dimensions, we focus primarily on two-dimensional CFTs and their defect CFTs here.

An important feature of this construction is that, unlike the analytic continuation of the two-dimensional Potts and $O(n)$ models, the central charge remains unchanged. Consequently, all theories on the complexified conformal manifold intrinsically belong to the same theory space and are connected by exactly marginal deformations. In other words, our construction realizes complex CFTs at a fixed central charge (which can take real values), where conformal invariance is preserved while scaling dimensions and other physical data become genuinely complex.
Moreover, the construction is completely general. Whenever a CFT, whether unitary or not, contains an appropriate class of exactly marginal operators, one can naturally obtain a continuous family of complex CFTs through analytic continuation of the corresponding couplings. This places complex fixed points inside ordinary conformal manifolds rather than treating them as isolated objects that appear only after analytically continuing parameters of the underlying theory space.

A central question is whether these analytically continued quantities are merely formal,
or whether they control universal observables in microscopic non-Hermitian critical
systems.  We show that the latter can occur: complex scaling dimensions, effective
central charges, and transmission coefficients extracted from the continued CFT agree
with biorthogonal observables in solvable lattice realizations.

Taken together, our results establish that starting from a unitary CFT with an exactly marginal bulk, boundary, or defect operator, one can naturally generate continuous families of nonunitary complex CFTs while preserving the original central charge. The resulting complex theories remain embedded in the same theory space as their unitary counterparts and possess analytically tractable conformal data. They therefore provide explicit and controllable platforms for investigating the broader landscape of complex CFTs and their applications to nonunitary critical phenomena.

%%%%%%%%%%%%%%%%%%%%%%%%%%%%%%%%%%%%%%%%%%%%%%%%%%%%%%%%%%%%%%%%%%%%%%%%%%%%%%%%%%%%%%%%%%%%%%
\subsection{Summary}
%%%%%%%%%%%%%%%%%%%%%%%%%%%%%%%%%%%%%%%%%%%%%%%%%%%%%%%%%%%%%%%%%%%%%%%%%%%%%%%%%%%%%%%%%%%%%%

In Sec.~\ref{sec:setup}, we study the compact free boson as the simplest bulk example.
Complexifying the Gaussian modulus, $R^2\in\mathbb{C}\setminus\{0\}$, gives an exactly solvable complex conformal manifold.
We determine its convergence domain and real locus.
This solvable example also lets us ask an important question: does a complex
conformal manifold contain new {\it rational points} away from the unitary locus?  We use the
compact boson as a diagnostic and then formulate the general obstruction.
We also explain that higher-rank Gaussian theories can nevertheless have
``semi-rational'' loci, where only part of the chiral algebra is rationally extended.

In Sec.~\ref{sec:IsingDefects}, we turn to interface CFTs, where the physical interpretation of complex
conformal data can be tested more directly.
Our main analytic example is the
one-parameter family of conformal defects in the Ising CFT, which is parametrized by an angular modulus $\phi_0$
\cite{Oshikawa1996,Oshikawa:1996ww}.
We analytically continue this modulus to $\phi_0\in\mathbb{C}$.
The defect operator spectrum is
\begin{equation}
\Delta_n(\phi_0)
=
2\left(n+\frac{\phi_0}{\pi}\right)^2,
\qquad n\in\mathbb{Z},
\label{eq:intro_defect_spectrum}
\end{equation}
so the defect two-point functions acquire complex powers away from the unitary slice.
This gives a direct defect-CFT analogue of the log-periodic scaling found in the bulk
compact boson.

A second important datum is the energy transmission coefficient of the conformal
interface \cite{Quella2006}.  On the unitary defect family in the Ising CFT, the reflection and transmission
coefficients are
\begin{equation}
\mathcal{R}(\phi_0)=\cos^2(2\phi_0),
\qquad
\mathcal{T}(\phi_0)=\sin^2(2\phi_0),
\qquad
\mathcal{R}+\mathcal{T}=1 .
\label{eq:intro_defect_transmission}
\end{equation}
For real $\phi_0$ these are ordinary non-negative energy reflection and transmission coefficients.  More generally,
they can be characterized by stress-tensor two-point data across the interface.  In
particular, for an interface between two CFTs with stress tensors $T_L$ and $T_R$, the
coefficient $c_{LR}$ in
\begin{equation}
\langle T_L(z_1)T_R(z_2)\rangle_I
=
\frac{c_{LR}/2}{(z_1-z_2)^4}
\end{equation}
determines the energy transmission coefficients in the unitary theory \cite{Meineri2019}.
After analytic continuation to complex $\phi_0$, the same formula defines a complex
coefficient of the stress-tensor two-point function.  It should
not be interpreted as a probability away from the unitary slice, although the
identity $\mathcal{R}+\mathcal{T}=1$ continues to hold.

A third observable is the entanglement entropy across an interface, which has a logarithmic term governed by an
interface effective central charge $c_{\rm eff}$ \cite{Sakai:2008tt,Brehm:2015lja}.
The scaling form is
\begin{equation}
S=\frac{c_{\rm eff}}{6}\log L+O(1).
\label{eq:intro_entanglement_scaling}
\end{equation}
The complexified effective central charge is obtained by analytically
continuing this function with a specified choice of sheet.  The absolute value appearing
in the real unitary expression is therefore replaced by a branch choice.
In particular, the factorizing defect is a logarithmic branch point of
$c_{\rm eff}$, while the topological defect is regular on the principal sheet.

In Sec.~\ref{sec:lattice}, these analytic predictions are tested in microscopic non-Hermitian
lattice models.  First, we study the critical transverse-field Ising chain with a complex
central bond.
For complex bond strength, the Hamiltonian is non-Hermitian.  We therefore use left and
right ground states and compute biorthogonal expectation values.  Bulk-defect
correlators extract the complex scaling dimension in the spin defect sector, while the
von Neumann entropy of the biorthogonal reduced density matrix extracts
$c_{\rm eff}$ through Eq.~\eqref{eq:intro_entanglement_scaling}.  Both quantities agree
with the analytic continuation of the Ising defect CFT predictions, including their real
and imaginary parts.
Second, we study a closely related free Dirac chain with a complex conformal interface.
A local joining quench creates an energy wave packet
that scatters off the interface.  In a biorthogonal time-evolved state, the expectation
value of the local Hamiltonian density is generally complex.  By integrating the
incoming and transmitted wave packets, we obtain a complex flux ratio.  This ratio agrees
with the analytic continuation of the unitary Dirac-interface result.
Thus the analytically continued transmission coefficient is operationally measurable as
a ratio of complex biorthogonal energy fluxes.  It is not a positive probability, but it
is a universal interface datum.

In Sec.~\ref{sec:ComplexBoundaryRGFlows}, we use AdS/BCFT to address a question that is not visible on an
exactly marginal deformation.  The bulk and defect examples above show
that conformal data can be analytically continued while conformal invariance is
preserved.  Since every point on such a manifold is already a fixed point,
however, these examples do not test any ordering principle along RG flow.
Boundary RG flows provide the natural next arena: after complexification, the
positivity assumptions behind monotonicity theorems can fail.  AdS/BCFT makes
this failure especially transparent, because the relevant positivity condition
appears as a sign in the classical equation for the end-of-the-world (EOW)
brane.
For real brane matter, the null-energy term has the sign that gives the
holographic boundary $g$-theorem.  Continuing a brane scalar to the imaginary
real-saddle branch, $\phi=i\chi$, reverses this sign.  The EOW-brane embedding
and $\log g$ can nevertheless remain real, so the violation is not an artifact
of comparing complex numbers.  One obtains a real saddle with
\begin{equation}
\log g_{\rm IR}>\log g_{\rm UV}.
\end{equation}
The same framework also has an interesting consequence.  At a boundary critical
point, the brane potential defines an effective tension $T_\ast$.  Ordinary AdS/BCFT fixed points are subcritical, $|T_\ast|L<1$, and have
AdS$_2$ branes.  An imaginary critical point of a complex boundary potential can
instead be supercritical, $|T_\ast|L>1$, in which case the classical AdS/BCFT equations admit a de Sitter brane.
This suggests that dS branes can be analyzed by considering complex relevant deformations.

%%%%%%%%%%%%%%%%%%%%%%%%%%%%%%%%%%%%%%%%%%%%%%%%%%%%%%%%%%%%%%%%%%%%%%%%%%%%%%%%
%%%%%%%%%%%%%%%%%%%%%%%%%%%%%%%%%%%%%%%%%%%%%%%%%%%%%%%%%%%%%%%%%%%%%%%%%%%%%%%%
\section{Complex Bulk Conformal Manifold}
\label{sec:setup}
%%%%%%%%%%%%%%%%%%%%%%%%%%%%%%%%%%%%%%%%%%%%%%%%%%%%%%%%%%%%%%%%%%%%%%%%%%%%%%%%
%%%%%%%%%%%%%%%%%%%%%%%%%%%%%%%%%%%%%%%%%%%%%%%%%%%%%%%%%%%%%%%%%%%%%%%%%%%%%%%%

We first discuss the simplest bulk example: the compact free boson with a complexified radius modulus.
This model is useful because every relevant CFT datum can be written explicitly and analytically continued without relying on perturbation theory.
It has also appeared as an effective description of dissipative Tomonaga--Luttinger liquids, for example in Ref.~\cite{Yamamoto2021}.
Our purpose here is to isolate the structural consequences of complexifying an exactly marginal bulk coupling.
We will define the continued theory through analytic continuation of standard CFT data, describe its spectrum and correlators, determine the convergence domain of the torus partition function, and identify distinguished loci such as the real locus.

%%%%%%%%%%%%%%%%%%%%%%%%%%%%%%%%%%%%%%%%%%%%%%%%%%%%%%%%%%%%%%%%%%%%%%%%%%%%%%%%
\subsection{Complex exactly marginal deformation in the free boson CFT}
\label{sec:setup:real}
%%%%%%%%%%%%%%%%%%%%%%%%%%%%%%%%%%%%%%%%%%%%%%%%%%%%%%%%%%%%%%%%%%%%%%%%%%%%%%%%

We consider the $c=1$ compact boson with field $X\sim X+2\pi R$, whose action is
\begin{equation}
S=\frac{1}{4\pi}\int d^2z\,\partial X\,\bar\partial X.
\end{equation}
This theory admits a continuous deformation generated by the exactly marginal operator 
$\del X \bar{\del} X$, 
which shifts the compactification radius.
It is convenient to factor out the periodicity by writing $X=R\,\theta$, where
\begin{equation}
\theta \sim \theta+2\pi,
\end{equation}
so that the radius dependence is carried by the coupling $R^2$,
\begin{equation}
S=\frac{R^2}{4\pi}\int d^2z\,\partial\theta\,\bar\partial\theta.
\label{eq:GaussianAction_r}
\end{equation}

We now complexify the exactly marginal coupling,
\begin{equation}
R^2 \in \mathbb{C}\setminus\{0\},
\label{eq:rComplex_rewrite}
\end{equation}
and define a family of \emph{complex} CFTs by analytic continuation from the standard real axis.
Concretely, for any observable whose expression is known for $R^2>0$,
we \emph{define} its value at complex $R^2$ to be the analytic continuation of that expression.
The beta function for the radius modulus vanishes on the real conformal manifold,
$\beta(R^2)=0$ for $R^2>0$; by analyticity, the same vanishing holds after complexifying
$R^2$ within the analytic domain, so the continued theory remains conformal although it is
in general nonunitary.

The fundamental operators in the compact free boson CFT are the vertex operators given by
\begin{equation}
V_{n,w}(z,\bar z)=:\ex{ip_LX_L(z)+ip_RX_R(\bar{z})}:
\label{eq:Vertex}
\end{equation}
where $X=X_L(z)+X_R(\bar z)$ and the left/right charges are defined by
\begin{equation}
p_L=\frac{n}{R}+wR,\qquad
p_R=\frac{n}{R}-wR, \qquad
n,w \in \bb{Z}.
\label{eq:pLpR}
\end{equation}
The conformal dimensions of $V_{n,w}$ are
\begin{equation}
(h_{n,w}, \bar h_{n,w}) = \pa{\frac{p_L^2}{4}, \frac{p_R^2}{4}},
\label{eq:hbarh}
\end{equation}
so that the scaling dimension and spin are
\begin{align}
\Delta_{n,w} &= h+\bar h
= \frac{1}{2}\left(\frac{n^2}{R^2}+w^2 R^2\right),
\label{eq:DeltaNW}\\
s_{n,w} &= h-\bar h
= n w \in\mathbb{Z}.
\label{eq:SpinNW}
\end{align}
For complex $R^2$, the weights are generically complex, implying that correlation functions
exhibit power laws controlled by $\mathrm{Re}\,\Delta$ dressed by log-periodic phases controlled by
$\mathrm{Im}\,\Delta$.
At the same time, the \emph{spin quantization}~\eqref{eq:SpinNW} implies that bulk correlators are
single-valued under $z\to \ex{2\pi i}z$, a key consistency property that survives complexification.

A closely related statement concerns mutual locality in the OPE.
For two vertex operators with labels $(n,w)$ and $(n',w')$, the monodromy exponent is governed by
\begin{equation}
\frac{1}{2}\bigl(p_L p_L' - p_R p_R'\bigr)=n w' + n' w \in \mathbb{Z},
\label{eq:MutualLocality}
\end{equation}
which is independent of $R^2$.
Thus the compactification lattice automatically enforces mutual locality even when the scaling
dimensions become complex.

\subsection{Correlation functions and log-periodic scaling}
\label{sec:setup:correlators}

The two-point function of vertex operators takes the standard form
\begin{equation}
\langle V_{n,w}(z,\bar z)\,V_{-n,-w}(0)\rangle
= z^{-2h_{n,w}}\,\bar z^{-2\bar h_{n,w}}
=|z|^{-2\Delta_{n,w}}\,\ex{-2is_{n,w}\arg z}.
\label{eq:TwoPoint}
\end{equation}
Writing $\Delta=\Delta_R+i\Delta_I$ with $\Delta_R, \Delta_I \in \bb{R}$, the radial dependence becomes
\begin{equation}
|z|^{-2\Delta}=|z|^{-2\Delta_R} \ex{-2i\Delta_I\log|z|},
\label{eq:LogPeriodic}
\end{equation}
exhibiting log-periodic oscillations whenever $\Delta_I\neq0$.
This behavior is the correlator analogue of the log-periodic scaling induced by complex critical
exponents in complexified RG flows and complex fixed points \cite{Tang:2024blm,Linden:2025nfc,Yamamoto2026}.

%%%%%%%%%%%%%%%%%%%%%%%%%%%%%%%%%%%%%%%%%%%%%%%%%%%%%%%%%%%%%%%%%%%%%%%%%%%%%%%%%%%%%%%%%%%%%%
\subsection{Torus partition function}
\label{sec:setup:torus}
%%%%%%%%%%%%%%%%%%%%%%%%%%%%%%%%%%%%%%%%%%%%%%%%%%%%%%%%%%%%%%%%%%%%%%%%%%%%%%%%%%%%%%%%%%%%%%

The torus partition function is
\begin{equation}
Z(\tau;R)=\frac{1}{|\eta(\tau)|^2}\sum_{n,w\in\mathbb{Z}}
q^{\fr{p_L^2}{4}}\,\bar q^{\fr{p_R^2}{4}},\qquad q=\ex{2\pi i\tau}.
\label{eq:ZtauR}
\end{equation}
Using $p_L^2+p_R^2=2(n^2/R^2+w^2 R^2)$ and $p_L^2-p_R^2=4nw$, one can rewrite~\eqref{eq:ZtauR} as
\begin{equation}
Z(\tau;R)=\frac{1}{|\eta(\tau)|^2}\sum_{n,w\in\mathbb{Z}}
\exp\!\left[-\pi\tau_2\left(\frac{n^2}{R^2}+w^2 R^2\right)+2\pi i\tau_1 n w\right].
\label{eq:ZexpForm}
\end{equation}
Absolute convergence of the lattice sum is controlled by the real part of the quadratic form.
For $R^2=a+ib$ with $a,b \in \bb{R}$, we have
\begin{equation}
\mathrm{Re}\left(\frac{n^2}{R^2}+w^2 R^2\right)
= \frac{a}{a^2+b^2}\,n^2+a\,w^2.
\label{eq:ReQuadraticForm}
\end{equation}
Hence for $\mathrm{Re}\,R^2=a>0$, the exponent in~\eqref{eq:ZexpForm} is strictly damped at large $|n|$
and $|w|$, ensuring absolute convergence for any $\tau_2>0$.
On the boundary $\mathrm{Re}\,R^2=0$, the damping disappears and the sum ceases to converge absolutely,
while for $\mathrm{Re}\,R^2<0$, it diverges exponentially.
This ``stability'' condition has been pointed out in \cite{Yamamoto2021}.

The partition function is manifestly invariant under the T-duality transformation
\begin{equation}
R \longmapsto \frac{1}{R},\qquad (n,w)\longmapsto (w,n),
\label{eq:Tduality}
\end{equation}
which preserves~\eqref{eq:ZexpForm} by relabeling the summation variables.
Modular invariance follows from the usual Poisson-resummation argument in the convergent domain $\mathrm{Re}\,R^2>0$ and then extends by analytic continuation within that domain.

It is convenient to parametrize the compactification radius by
\begin{equation}
R=\ell\ex{i\phi}.
\label{eq:tlogr}
\end{equation}
The partition function converges absolutely whenever
$\mathrm{Re}\,R^2>0$.
In the polar coordinates, this becomes
\begin{equation}
\mathrm{Re}\,R^2>0
\quad\Longleftrightarrow\quad
\phi\in\Bigl(-\frac{\pi}{4},\frac{\pi}{4}\Bigr),
\label{eq:bulkstrip}
\end{equation}
where we quotient by $R\sim -R$ and keep the wedge with $\mathrm{Re}\,R>0$.

%%%%%%%%%%%%%%%%%%%%%%%%%%%%%%%%%%%%%%%%%%%%%%%%%%%%%%%%%%%%%%%%%%%%%%%%%%%%%%%%%%%%%%%%%%%%%%
%%%%%%%%%%%%%%%%%%%%%%%%%%%%%%%%%%%%%%%%%%%%%%%%%%%%%%%%%%%%%%%%%%%%%%%%%%%%%%%%%%%%%%%%%%%%%%
\subsection{Distinguished regions on the complex Gaussian conformal manifold}
\label{sec:global}
%%%%%%%%%%%%%%%%%%%%%%%%%%%%%%%%%%%%%%%%%%%%%%%%%%%%%%%%%%%%%%%%%%%%%%%%%%%%%%%%%%%%%%%%%%%%%%
%%%%%%%%%%%%%%%%%%%%%%%%%%%%%%%%%%%%%%%%%%%%%%%%%%%%%%%%%%%%%%%%%%%%%%%%%%%%%%%%%%%%%%%%%%%%%%

\begin{figure}[t]
 \begin{center}
  \includegraphics[width=12.0cm,clip]{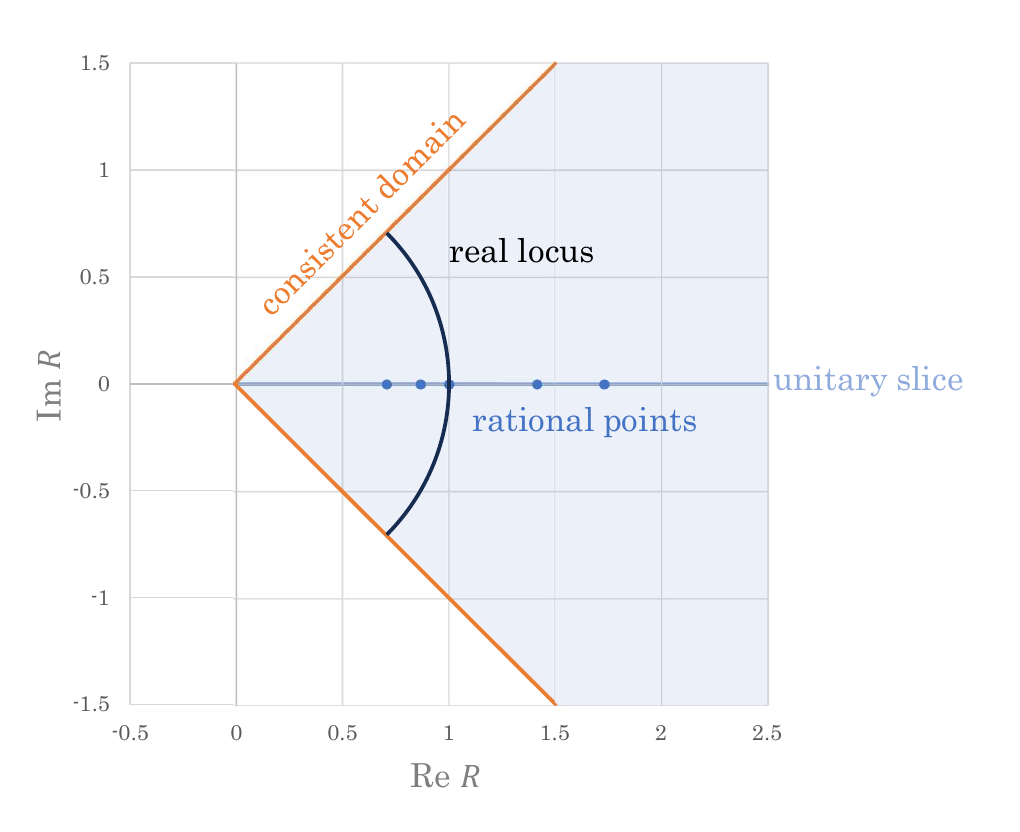}
 \end{center}
\caption{
Complex bulk conformal manifold in the $R$-plane.
The shaded wedge denotes the consistent domain $\mathrm{Re}(R^2)>0$, where the torus partition function converges absolutely.
The arc $|R|=1$ (the black line) inside this wedge is the real locus, on which complex conjugation combined with T-duality implies $Z(\tau;R)\in\mathbb{R}$.
Marked points on the positive real axis indicate rational loci with $R^2\in\mathbb{Q}_{> 0}$.
}
 \label{fig:CCM}
\end{figure}

The complex Gaussian family constructed in Sec.~\ref{sec:setup} is an unusual instance of a \emph{complex} CFT that is conformal by construction (beta functions vanish identically along the manifold), while still exhibiting genuinely complex universal data (complex scaling dimensions and log-periodic correlators).
Below, we show that the resulting complex conformal manifold has a rich global structure, and we identify several sharply defined, physically distinguished regions and loci, which are summarized in Figure \ref{fig:CCM}.

%%%%%%%%%%%%%%%%%%%%%%%%%%%%%%%%%%%%%%%%%%%%%%%%%%%%%%%%%%%%%%%%%%%%%%%%%%%%%%%%%%%%%%%%%%%%%%
\subsubsection{Real locus of complex CFT data}
\label{sec:global:dualitywall}
%%%%%%%%%%%%%%%%%%%%%%%%%%%%%%%%%%%%%%%%%%%%%%%%%%%%%%%%%%%%%%%%%%%%%%%%%%%%%%%%%%%%%%%%%%%%%%

A particularly striking feature of the complex Gaussian manifold is the existence of a codimension-one
locus on which important CFT data obey a \emph{reality condition} despite the theory being nonunitary.

First, complex conjugation acts on the torus partition function as
\begin{equation}
Z(\tau;R)^\ast = Z(\tau;R^\ast).
\label{eq:Z_conj}
\end{equation}
Second, the theory is exactly invariant under T-duality,
\begin{equation}
Z(\tau;R)=Z(\tau;1/R),
\label{eq:Z_duality}
\end{equation}
by the relabeling $(n,w)\leftrightarrow(w,n)$.
Combining~\eqref{eq:Z_conj} and~\eqref{eq:Z_duality}, we find that whenever
\begin{equation}
R^\ast=\frac{1}{R}\qquad\Longleftrightarrow\qquad |R|=1,
\label{eq:real_locus_condition}
\end{equation}
the partition function is real:
\begin{equation}
|R|=1 \ \ \Rightarrow\ \ Z(\tau;R)^\ast=Z(\tau;1/R)=Z(\tau;R).
\label{eq:Z_real_on_wall}
\end{equation}
Thus the arc $|R|=1$ inside $\mathrm{Re}\,R^2>0$ defines a \emph{real locus} of the complex Gaussian
manifold where the torus partition function is guaranteed to be real by symmetry.

The same mechanism organizes the spectrum.
On $|R|=1$ we have
\begin{equation}
\Delta_{n,w}(R)^\ast=\Delta_{n,w}(R^\ast)=\Delta_{n,w}(1/R)=\Delta_{w,n}(R),
\label{eq:spectrum_pairing}
\end{equation}
so complex scaling dimensions pair into conjugate partners under $(n,w)\leftrightarrow (w,n)$.
This spectral pairing is the cleanest manifestation, in a fully solvable setting, of a general theme
in complexified field theories: complex universal data often organize into conjugate multiplets and
can enforce real bulk observables even in nonunitary theories.

%%%%%%%%%%%%%%%%%%%%%%%%%%%%%%%%%%%%%%%%%%%%%%%%%%%%%%%%%%%%%%%%%%%%%%%%%%%%%%%%%%%%%%%%%%%%%%
\subsubsection{Enhanced symmetry loci and their persistence under complexification}
\label{sec:global:enhanced}
%%%%%%%%%%%%%%%%%%%%%%%%%%%%%%%%%%%%%%%%%%%%%%%%%%%%%%%%%%%%%%%%%%%%%%%%%%%%%%%%%%%%%%%%%%%%%%
On the real conformal manifold, special rational values of $R^2$ produce enhanced chiral symmetry
(e.g., current algebra extensions and rational CFT points).
Complexification does \emph{not} generically proliferate such points.
The reason is simple and robust: enhanced chiral symmetries require chiral operators, which impose arithmetic conditions such as $p_R=0$ or $p_L=0$.
For the compact boson, these conditions reduce to $R^2\in\mathbb{Q}\subset\mathbb{R}$,
so enhanced symmetry remains confined to the real axis even after complexifying $R^2$.
Thus complexification generates new global structures (consistent wedge, real locus) without creating new islands of rationality.

%%%%%%%%%%%%%%%%%%%%%%%%%%%%%%%%%%%%%%%%%%%%%%%%%%%%%%%%%%%%%%%%%%%%%%%%%%%%%%%%%%%%%%%%%%%%%%
\subsubsection{Convergence boundary and stability}
\label{sec:global:instability}
%%%%%%%%%%%%%%%%%%%%%%%%%%%%%%%%%%%%%%%%%%%%%%%%%%%%%%%%%%%%%%%%%%%%%%%%%%%%%%%%%%%%%%%%%%%%%%

The condition $\mathrm{Re}\,R^2>0$ should not be confused with conformality: the beta function still
vanishes after analytic continuation wherever the CFT data are analytic.
Rather, it is a stability condition for the Euclidean torus sum.
Inside the wedge the momentum and winding sums are exponentially damped; on the boundary
$\mathrm{Re}\,R^2=0$ the damping is lost; outside the wedge the lattice sum is exponentially
amplified.
Outside $\mathrm{Re}\,R^2>0$, we only claim a formal analytic continuation of local conformal data. We do not claim the existence of a convergent Euclidean path integral or canonical torus trace.

%%%%%%%%%%%%%%%%%%%%%%%%%%%%%%%%%%%%%%%%%%%%%%%%%%%%%%%%%%%%%%%%%%%%%%%%%%%%%%%%%%%%%%%%%%%%%%
%%%%%%%%%%%%%%%%%%%%%%%%%%%%%%%%%%%%%%%%%%%%%%%%%%%%%%%%%%%%%%%%%%%%%%%%%%%%%%%%%%%%%%%%%%%%%%
\subsection{Rational points on complex manifolds?}
\label{sec:rational}
%%%%%%%%%%%%%%%%%%%%%%%%%%%%%%%%%%%%%%%%%%%%%%%%%%%%%%%%%%%%%%%%%%%%%%%%%%%%%%%%%%%%%%%%%%%%%%
%%%%%%%%%%%%%%%%%%%%%%%%%%%%%%%%%%%%%%%%%%%%%%%%%%%%%%%%%%%%%%%%%%%%%%%%%%%%%%%%%%%%%%%%%%%%%%
Next, let us consider special points scattered throughout the moduli space of complex CFTs. There is some freedom in deciding what should be regarded as special, but one natural idea is to extend to the complex domain the property familiar for real compact bosons, namely that the partition function can be decomposed into a finite sum of characters of highest-weight representations of some chiral algebra. We will call a point with this property a rational point. 

With rationality defined in this way, one may ask whether rationality can occur at a general point in the complex conformal manifold.
As stated in Sec.~\ref{sec:global:enhanced}, there is no rational point in the complex conformal manifold for $c=1$ compact boson.
One may then ask whether, in a general CFT, there exist rational points. The answer is again negative. In fact, this follows from a result that has been known for a long time. Vafa showed that, once one imposes rationality, in the sense that there are only finitely many primary fields, one can derive a set of equations known as Vafa’s equations \cite{Vafa:1988ag,Anderson:1987ge}. By using the transformation properties under modular transformations in these equations, one can show that $c,h$ must be rational numbers\footnote{For mathematical references, see\cite{Etingof2002OnVT,miyamoto2004modular}. In particular, the paper by Miyamoto \cite{miyamoto2004modular} discusses the result in a purely mathematical way, starting from the assumption that the VOA is $\text{C}_2$-cofinite, and the proof seems to have nothing to do with Vafa's equation.}.

Roughly speaking, these equations are equations for the “twists” $\alpha_i=e^{2\pi i h_i}$ associated with the primary fields and the fusion coefficients $N_{ijk}$, where $h_i$ denotes the conformal dimension of the primary field $i$. Let us first introduce $N_{ijkl}$, the dimension of the space of four-point conformal blocks. Then, by considering a special case of Vafa’s equations, one obtains
\begin{equation}
    (\alpha_i)^{4N_{iiii}-3N_{iii}}\prod_{r\neq i}(\alpha_r)^{-3N_{iir}}=1.
\end{equation}
If rationality is imposed, the set of primary-field labels is finite. We may therefore define a finite-dimensional integer matrix $M$ by
\begin{equation}
    M_{ir}=\delta_{r,i}(4N_{iiii}-3N_{iii})+(1-\delta_{r,i})(-3N_{iir}),
\end{equation}
and introduce the vector $\mathbf{h}$ by $\mathbf{h}_i=h_i$. The above equation is then equivalent to
\begin{equation}
    M\mathbf{h}\in \mathbb{Z}^{n}.
\end{equation}
In fact, a simple argument shows that $M$ is invertible. Since all entries of $M$ are integers, it follows that each component of $\mathbf{h}$ must be rational. Thus all conformal dimensions are rational.
The rationality of the central charge follows from the properties of the modular $T$ matrix. As explained, for example, in Etingof’s paper\cite{Etingof2002OnVT}, the relations $(ST)^3=S^4=1$ imply that $\det T$ is a root of unity. On the other hand, the modular $T$ matrix acts on the characters of the primary fields by phases of the form $e^{2\pi i (h_i-c/24)}$. Since all $h_i$ have already been shown to be rational, the fact that $\det T$ is a root of unity implies that $c$ must also be rational.

An important point is that this argument does not assume from the outset that $c,h$ are real numbers. The same reasoning applies to CFTs whose parameters, such as the central charge, may take general complex values. Thus, one concludes that complex CFTs do not admit rational points. On the other hand, as we will discuss in Sec.~\ref{sec:Discussion}, when one considers CFTs with boundaries, there are cases in which the boundary spectrum admits a finite decomposition, thereby acquiring a certain sense of rationality.

\paragraph{Semi-rational points.}

Does it then follow that, even for Narain CFTs with the central charge greater than 1, no nontrivial and interesting theories appear in the complex domain? The answer is again no. We can see that there are points that can be called \emph{semi-rational} points.
Let us first give an example in the case $c=2$ \footnote{For more general case $c\geq 3$, one can easily extend the following discussion, and state that there are continuous families of semi-rational points.}where, by choosing $E:= G+B$ appropriately, the lattice generated by chiral primaries has rank 1, and hence the partition function can be decomposed to some extent. Let us choose

\begin{equation}
    G=\begin{pmatrix}1 & \frac{i}{2} \\ \frac{i}{2} & 1\end{pmatrix},B=\begin{pmatrix}0& \frac{i}{2} \\ -\frac{i}{2} & 0\end{pmatrix}.
\end{equation}
Then we have
\begin{equation}
    \frac{p_L^2}{2}=\frac{1}{4}((G+B)N+M)^{\top}G^{-1}((G+B)N+M)
=\frac{1}{4}\left| \left|\begin{pmatrix} m_1+n_1+in_2 \\ m_2+n_2\end{pmatrix}\right| \right|^2_{G^{-1}}
\end{equation}
where $||\cdot ||_{A}^2$ denotes the square of a vector defined by the complex quadratic form $A$. Now, introducing the change of variables

\begin{equation}
    a=m_1+n_1, b=m_2-n_2, k=m_2, t=n_1
\end{equation}
and computing $p_R$ as well, we obtain

\begin{equation}
    \begin{split}
        \frac{p_L^2}{2}&=\frac{1}{4}\left| \left|\begin{pmatrix}a-ib\\ -b\end{pmatrix}+k\begin{pmatrix}i \\ 2\end{pmatrix}\right|\right|^2_{G^{-1}}=\frac{1}{4}||\lambda_{a,b}+k l_1||^2_{G^{-1}}\\
\frac{p_R^2}{2}&=\frac{1}{4}\left|\left|\begin{pmatrix}-a \\ -b\end{pmatrix}+t\begin{pmatrix}2 \\ i \end{pmatrix}\right|\right|^2_{G^{-1}}=\frac{1}{4}||\mu_{a,b}+tl_2||^2_{G^{-1}}
    \end{split}
\end{equation}

The important point is that, when summing over these contributions, the variables $k$ and $t$ are independent. Therefore, once $a,b$ are fixed, the sum of $q^{h_L}\bar{q}^{h_R}$ factorizes into a purely left-moving sum and a purely right-moving sum. That is,

\begin{equation}
    \mathcal{Z}(\tau)=\frac{1}{|\eta|^4}\sum_{p_L,p_R}q^{\frac{p_L^2}{2}}\bar{q}^{\frac{p_R^2}{2}}=
\sum_{a,b\in\mathbb{Z}}\chi_{a,b}^{L}(\tau)\chi_{a,b}^{R}(\bar{\tau})
\end{equation}
where
\begin{equation}
    \begin{split}
        \chi_{a,b}^{L}(\tau)&=\frac{1}{\eta^2}\sum_k q^{\frac{1}{4}||\lambda_{a,b}+k l_1||^2_{G^{-1}}}\\
\chi_{a,b}^{R}(\bar{\tau})&=\frac{1}{\bar{\eta}^2}\sum_t \bar{q}^{\frac{1}{4}|| \mu_{a,b}+t l_2||^2_{G^{-1}}}.
    \end{split}
\end{equation}

Furthermore, these characters can be rewritten. For example,

\begin{equation}
    \chi_{a,b}^{L}(\tau)=\frac{1}{\eta}q^{\frac{1}{5}(a-\frac{i}{2}b)^2}\frac{1}{\eta}\sum_k q^{(k-\frac{b}{2})^2}
\end{equation}
which takes the form of a product of the character of a Heisenberg VOA, with complexified weight, and the character of the $\widehat{\mathfrak{su}(2)}_1$ WZW model. This may be regarded as the representation of the Deligne product ``$\boxtimes$" of two chiral algebras, where one of the two U(1) factors is “rational” while the other is the chiral algebra of a “complexified” boson. In this sense, it should be viewed as a theory that is  \emph{semi-rational}. This can be understood as performing a simple current extension along a rank-1 sublattice inside the lattice generated by $p_L$. Moreover, such an extension is possible provided that there exist integer vectors $N^{L,R}_0$ satisfying $(G+B)N^L_0\in\mathbb{Z}^2,(G-B)N_0^R\in\mathbb{Z}^2$.
This condition is equivalent, in terms of $E:=G+B$, to $EN^L_0,\, (N_0^R)^{\top}E\in\mathbb{Z}^2$. How many such $E$ exist, and how are they distributed? The answer is that they are given by
\begin{equation}
    E=U\begin{pmatrix} a & b \\ c & z \end{pmatrix}V,\qquad U,V\in\mathrm{SL}(2,\mathbb{Z})
\end{equation}
where $a,b,c\in\mathbb{Q}$, $z\in\mathbb{C}$. In other words, such points are distributed along lines with a one-complex-dimensional line, which is continuous in the complex manifold. This is in sharp contrast to rational points, which are dense in the real moduli space.

It is useful to give a physical interpretation of the semi-rational loci described above. A semi-rational point should not be regarded as a rational CFT in the usual sense: the full theory still contains an infinite set of sectors and generically has complex conformal weights. Rather, it is a point on the complexified Gaussian conformal manifold at which only a sublattice of the momentum-winding lattice supports an extended chiral algebra. Equivalently, only part of the gapless theory is organized by a rational chiral algebra, while the remaining part is a genuinely complex, non-rational Gaussian sector.

From this viewpoint, semi-rationality means that the complex CFT contains a protected rational subsector. This has a direct physical meaning in multi-component Luttinger liquids\cite{Haldane1981}. In an ordinary spinful or SU($N$)-symmetric Luttinger liquid, the low-energy theory often separates into a charge sector and a spin sector. If the system is made dissipative or non-Hermitian in such a way that the charge sector acquires a complex Luttinger parameter, while the spin symmetry remains intact, the infrared theory naturally takes the schematic form
\[
\text{complex } U(1) _{\text{ Gaussian sector}}
\quad \boxtimes \quad
\text{rational spin sector}.
\]
The rational factor is not an accidental finite-dimensional truncation of the full Hilbert space. It is the remnant of a microscopic symmetry that fixes part of the infrared conformal data. Consequently, correlation functions may exhibit complex scaling dimensions and log-periodic behavior through the charge sector, while their spin quantum numbers, selection rules, and part of their conformal towers are still controlled by a finite rational chiral algebra.

A closely related physical realization appears in the dissipative Tomonaga--Luttinger liquids with SU($N$) spin symmetry studied by Yamamoto and Kawakami \cite{YamamotoKawakami}. They found that the low-energy spectrum is described by the sum of one charge mode, governed by a complex generalization of the $c=1$ U(1) Gaussian CFT, and ($N-1$) spin modes governed by the level-one SU$(N)$ Kac--Moody algebra. In particular, dissipation affects the charge mode, whereas the spin sector remains described by the rational SU$(N)_1$ WZW theory as a consequence of spin-charge separation. This is precisely the physical mechanism that our semi-rational loci abstractly capture: non-Hermitian or dissipative effects complexify only part of the Gaussian conformal data, while a symmetry-protected rational sector survives.

Thus semi-rational points provide a useful bridge between complex conformal manifolds and experimentally or numerically motivated non-Hermitian one-dimensional systems. They identify loci where the full theory is not rational, but where part of the operator algebra, fusion rules, and finite-size spectrum remain algebraically controlled. In practical terms, this partial rationality can constrain the classification of operators, organize finite-size spectra, and separate the origin of complex critical exponents from symmetry-protected rational data. In this sense, semi-rationality is not merely a formal arithmetic property of the complex Narain moduli space, but a field-theoretic avatar of spin-charge separated dissipative Luttinger liquids.

%%%%%%%%%%%%%%%%%%%%%%%%%%%%%%%%%%%%%%%%%%%%%%%%%%%%%%%%%%%%%%%%%%%%%%%%%%%%%%%%%%%%%%%%%%%%%%
%%%%%%%%%%%%%%%%%%%%%%%%%%%%%%%%%%%%%%%%%%%%%%%%%%%%%%%%%%%%%%%%%%%%%%%%%%%%%%%%%%%%%%%%%%%%%%
\section{Complex Defect Conformal Manifold}
\label{sec:IsingDefects}
%%%%%%%%%%%%%%%%%%%%%%%%%%%%%%%%%%%%%%%%%%%%%%%%%%%%%%%%%%%%%%%%%%%%%%%%%%%%%%%%

We now move from bulk conformal manifolds to defect conformal manifolds.
This is the setting in which complexification is both analytically tractable and directly testable in lattice models.
The Ising CFT provides a particularly useful example: its conformal defects form an exactly known one-parameter family, the folded description reduces the problem to boundary CFT in $\mathrm{Ising}\otimes\mathrm{Ising}$, and the same family has a simple realization as an energy defect in the critical transverse-field Ising chain.
The goal of this section is therefore twofold.
First, we analytically continue the exact defect CFT data---the defect spectrum, energy transmission coefficient, and entanglement coefficient---to a complex defect manifold.
Second, we organize the formulas in a way that makes contact with the lattice variables used in Sec.~\ref{sec:lattice}.

%%%%%%%%%%%%%%%%%%%%%%%%%%%%%%%%%%%%%%%%%%%%%%%%%%%%%%%%%%%%%%%%%%%%%%%%%%%%%%%%
\subsection{Real defect conformal manifold}
\label{sec:IsingDefects:real}
%%%%%%%%%%%%%%%%%%%%%%%%%%%%%%%%%%%%%%%%%%%%%%%%%%%%%%%%%%%%%%%%%%%%%%%%%%%%%%%%

The Ising CFT admits a one-parameter family of conformal defects generated by an exactly marginal defect perturbation,
\begin{equation}
S_{\rm Ising}
+
\lambda\int d\tau\,\varepsilon .
\label{eq:Ising_defect_lagrangian}
\end{equation}
Here $\varepsilon$ is the defect energy operator of scaling dimension one.
With this normalization the corresponding defect beta function vanishes along the real family,
\begin{equation}
\beta(\lambda)=0 .
\label{eq:beta_lambda_zero_defect}
\end{equation}

The finite coupling is conveniently encoded in a phase-shift coordinate $\alpha$,
\begin{equation}
\tan\frac{\alpha}{2}=\lambda,
\label{eq:lambda_alpha_relation}
\end{equation}
so that the transparent defect is at $\lambda=0$, or $\alpha=0$.
This local Lagrangian coordinate is related to the angular coordinate $\phi_0$ used in the exact boundary-state description by
\begin{equation}
\phi_0=\frac{\pi}{4}+\frac{\alpha}{2}
=
\frac{\pi}{4}+\arctan\lambda,
\label{eq:lambda_alpha_phi_relation}
\end{equation}
and hence
\begin{equation}
\lambda=\tan\left(\phi_0-\frac{\pi}{4}\right).
\label{eq:lambda_phi_relation}
\end{equation}
Thus $\lambda$ is a natural local coordinate near the transparent defect, while $\phi_0$ is a more global angular coordinate on the defect conformal manifold.
The pole of Eq.~\eqref{eq:lambda_phi_relation} at $\phi_0=3\pi/4$ is not a singularity of the defect CFT; it only means that the $\lambda$ patch does not cover the second topological defect.

The global description follows by folding.
A defect line in the Ising model becomes a conformal boundary condition in the doubled theory $\mathrm{Ising}\otimes\mathrm{Ising}$.
This doubled theory is equivalent to the $\mathbb{Z}_2$ orbifold of a compact boson, and the exactly marginal defect perturbation translates the corresponding orbifold Dirichlet brane.
In this language the family is the orbifold Dirichlet boundary state
\begin{equation}
\mathrm{D_O}(\phi_0),
\label{eq:orbifold_D_boundary_state}
\end{equation}
parameterized by the same angle $\phi_0$ \cite{Oshikawa1996}.
This folded description makes the full conformal manifold manifest and gives closed-form expressions for the partition function and defect operator spectrum.
Along the real exactly marginal family the defect entropy is unchanged, while transmission coefficients and defect scaling dimensions vary with $\phi_0$.

For the Ising defect family, the reflection and transmission coefficients are \cite{Quella2006}
\begin{equation}
\mathcal{R}(\phi_0)=\cos^2(2\phi_0),
\qquad
\mathcal{T}(\phi_0)=\sin^2(2\phi_0).
\label{eq:Ising_RT_real}
\end{equation}
Using Eqs.~\eqref{eq:lambda_alpha_relation} and \eqref{eq:lambda_alpha_phi_relation}, these can also be written in terms of the local Lagrangian coordinate as
\begin{equation}
\mathcal{T}(\lambda)
=
\cos^2\alpha
=
\left(\frac{1-\lambda^2}{1+\lambda^2}\right)^2,
\qquad
\mathcal{R}(\lambda)
=
\sin^2\alpha
=
\left(\frac{2\lambda}{1+\lambda^2}\right)^2 .
\label{eq:Ising_RT_lambda}
\end{equation}
For real $\lambda$, these are ordinary energy transmission coefficients and obey
\begin{equation}
0\leq \mathcal{R},\mathcal{T}\leq1,
\qquad
\mathcal{R}+\mathcal{T}=1.
\label{eq:Ising_RT_unitary_balance}
\end{equation}
The same algebraic formulas will be analytically continued below.

Several real points are especially important.
The transparent, or identity topological, defect is
\begin{equation}
\lambda=0
\qquad\Longleftrightarrow\qquad
\phi_0=\frac{\pi}{4},
\qquad
\mathcal{T}=1,
\label{eq:lambda_topological_identity}
\end{equation}
while the second topological defect is reached at the pole of the same $\lambda$ patch,
\begin{equation}
\lambda=\infty
\qquad\Longleftrightarrow\qquad
\phi_0=\frac{3\pi}{4},
\qquad
\mathcal{T}=1.
\label{eq:lambda_topological_second}
\end{equation}
The factorizing, totally reflecting points are
\begin{equation}
\lambda=\pm1
\qquad\Longleftrightarrow\qquad
\phi_0=0,\frac{\pi}{2}\quad (\mathrm{mod}\ \pi),
\qquad
\mathcal{T}=0.
\label{eq:lambda_factorizing}
\end{equation}

For later comparison with the lattice model, it is useful to introduce one more coordinate.
In Sec.~\ref{sec:lattice}, the Ising defect is represented by a central bond coupling $b_\epsilon$, with $b_\epsilon=1$ corresponding to the homogeneous transparent chain and $b_\epsilon=0$ to the cut chain.
The continuum angular coordinate is related to this lattice coordinate by
\begin{equation}
\phi_0(b_\epsilon)=\arctan\frac{1}{b_\epsilon},
\qquad
\lambda
=
\tan\left(\phi_0-\frac{\pi}{4}\right)
=
\frac{1-b_\epsilon}{1+b_\epsilon},
\label{eq:lattice_CFT_dictionary}
\end{equation}
where the branch is chosen so that $b_\epsilon=1$ gives $\phi_0=\pi/4$.
Consequently,
\begin{equation}
s:=\sin(2\phi_0)
=
\frac{2b_\epsilon}{1+b_\epsilon^2}.
\label{eq:s_lattice_b}
\end{equation}
This is the same variable denoted by $t$ in the numerical comparison of the effective central charge.

%%%%%%%%%%%%%%%%%%%%%%%%%%%%%%%%%%%%%%%%%%%%%%%%%%%%%%%%%%%%%%%%%%%%%%%%%%%%%%%%
\subsection{Complex defect conformal manifold}
\label{sec:IsingDefects:complexification}
%%%%%%%%%%%%%%%%%%%%%%%%%%%%%%%%%%%%%%%%%%%%%%%%%%%%%%%%%%%%%%%%%%%%%%%%%%%%%%%%

We now complexify the exactly marginal defect coupling.
In the Lagrangian coordinate this means
\begin{equation}
\lambda\in\mathbb{C},
\qquad
\phi_0=\frac{\pi}{4}+\arctan\lambda,
\label{eq:lambda_complex_phi0}
\end{equation}
with a specified local branch of $\arctan\lambda$.
Equivalently, and more globally, we analytically continue the angular coordinate itself,
\begin{equation}
\phi_0\in\mathbb{C}.
\label{eq:phi0_complex}
\end{equation}
Since the defect is conformal for all real $\phi_0$ and the relevant defect CFT data are analytic functions of $\phi_0$, we define the complex defect CFT by analytic continuation of these expressions.
In RG language, the defect beta function vanishes on the real defect conformal manifold; its analytic continuation therefore vanishes on the connected complex domain by the identity theorem.
The resulting theory preserves defect conformal symmetry even though unitarity is generically lost.

Let $L$ be the spatial circumference and $\beta$ the Euclidean time length.
Introduce
\begin{equation}
q=\ex{-2\pi\beta/L},
\qquad
z=\ex{-\alpha\,\beta/L},
\label{eq:q_w_def}
\end{equation}
where $\alpha$ is the defect phase shift.
The Ising partition function in the presence of the defect can be written as \cite{Oshikawa1996}
\begin{equation}
Z_{\rm Ising}(\alpha;\beta/L)
=
\frac{\ex{-\alpha^2\beta/(4\pi L)}}{2}
\left[
\frac{\vartheta_3(z,q)}{\eta(q)}+
\frac{\vartheta_4(z,q)}{\eta(q)}+
\frac{\vartheta_2(z,q)}{\eta(q)}+
 i\frac{\vartheta_1(z,q)}{\eta(q)}
\right],
\label{eq:IsingZ_theta}
\end{equation}
where $\vartheta_i$ and $\eta$ are the standard Jacobi and Dedekind functions.
For $|q|<1$, this expression is analytic in $z$, and hence its continuation to complex $\alpha$ is immediate.

The defect operator spectrum is \cite{Oshikawa1996}
\begin{equation}
\Delta_n(\phi_0)=2\left(n+\frac{\phi_0}{\pi}\right)^2,
\qquad n\in\mathbb{Z}.
\label{eq:defect_spectrum_moving}
\end{equation}
Writing $\phi_0/\pi=x+iy$, one obtains
\begin{equation}
\Delta_n(\phi_0)
=2\big[(n+x)^2-y^2\big]+i\,4y(n+x).
\label{eq:Delta_complex_components}
\end{equation}
Thus defect two-point functions behave as
\begin{equation}
\langle \ca{O}_n(\tau)\ca{O}_n(0)\rangle
\propto |\tau|^{-2\Delta_n}
=
|\tau|^{-2\,\mathrm{Re}\,\Delta_n}
\ex{-2i(\mathrm{Im}\,\Delta_n)\log|\tau|}.
\label{eq:defect_logperiodic}
\end{equation}
For $\mathrm{Im}\,\Delta_n\neq0$, the correlator exhibits log-periodic modulation along the defect, exactly as in bulk complex CFTs with complex critical exponents.

The lattice bulk-defect correlator studied in Sec.~\ref{sec:lattice} probes the branch continuously connected to the spin defect field at the transparent point.
With the branch choice in Eq.~\eqref{eq:lattice_CFT_dictionary}, the analytic prediction used in the numerical comparison is
\begin{equation}
\Delta^{\sigma}_{\rm defect}(b_\epsilon)
=
2\left(\frac{\phi_0(b_\epsilon)}{\pi}\right)^2,
\qquad
\phi_0(b_\epsilon)=\arctan\frac{1}{b_\epsilon},
\label{eq:Delta_sigma_defect_b}
\end{equation}
up to the integer shifts $\phi_0/\pi\to n+\phi_0/\pi$ that select other branches of the open-string spectrum.

%%%%%%%%%%%%%%%%%%%%%%%%%%%%%%%%%%%%%%%%%%%%%%%%%%%%%%%%%%%%%%%%%%%%%%%%%%%%%%%%
\subsection{Energy transmission coefficient}
\label{sec:IsingDefects:energytransmission}
%%%%%%%%%%%%%%%%%%%%%%%%%%%%%%%%%%%%%%%%%%%%%%%%%%%%%%%%%%%%%%%%%%%%%%%%%%%%%%%%

We next clarify the meaning of the analytically continued transmission coefficient.
On the unitary slice, a conformal interface can be probed by a collider experiment: one prepares a localized excitation far from the interface, lets it collide with the defect, and measures the energy flux reaching future null infinity.
In two dimensions these fluxes are measured by ANEC light-ray operators built from the stress tensors on the two sides of the interface, for example
\begin{equation}
\mathcal{E}_R=-\frac{1}{2\pi}\int dz\,T_R(z),
\qquad
\bar{\mathcal{E}}_L=-\frac{1}{2\pi}\int d\bar z\,\bar T_L(\bar z),
\label{eq:ANEC_defect_light_ray}
\end{equation}
with analogous definitions for $\mathcal{E}_L$ and $\bar{\mathcal{E}}_R$.
For a unitary interface, the corresponding transmission and reflection coefficients are genuine fractions of energy.
They are non-negative and obey $\mathcal{T}+\mathcal{R}=1$.

More generally, these coefficients are fixed by stress-tensor two-point data rather than by a particular wave packet.
Let $c_{LR}$ be the coefficient of the stress-tensor two-point function across the interface \cite{Meineri2019},
\begin{equation}
\langle T_L(z_1)T_R(z_2)\rangle_I
=
\frac{c_{LR}/2}{(z_1-z_2)^4}.
\label{eq:cLR_defect_def}
\end{equation}
Then
\begin{equation}
\mathcal{T}_L=\frac{c_{LR}}{c_L},
\qquad
\mathcal{T}_R=\frac{c_{LR}}{c_R}.
\label{eq:collider_transmission_general}
\end{equation}
For the Ising defect considered here, $c_L=c_R=c_{\rm Ising}=1/2$, so
\begin{equation}
\mathcal{T}(\phi_0)=\frac{c_{LR}(\phi_0)}{c_{\rm Ising}},
\qquad
c_{LR}(\phi_0)=\frac12\mathcal{T}(\phi_0)=\frac12\sin^2(2\phi_0).
\label{eq:cLR_transmission_Ising}
\end{equation}

The complex defect is obtained by continuing the same stress-tensor data to $\phi_0\in\mathbb{C}$.
Writing
\begin{equation}
\phi_0=u+i\eta,
\qquad u,\eta\in\mathbb{R},
\label{eq:phi0_u_eta}
\end{equation}
Eq.~\eqref{eq:Ising_RT_real} gives
\begin{align}
\mathcal{T}(\phi_0)
&=
\frac{1-\cos(4u)\cosh(4\eta)}{2}
+
\frac{i}{2}\sin(4u)\sinh(4\eta),
\label{eq:complex_T_explicit}
\\
\mathcal{R}(\phi_0)
&=
\frac{1+\cos(4u)\cosh(4\eta)}{2}
-
\frac{i}{2}\sin(4u)\sinh(4\eta).
\label{eq:complex_R_explicit}
\end{align}
In general $\mathcal{T}$ and $\mathcal{R}$ are complex.
They should not be interpreted as ordinary probabilities, because the positive Hilbert-space interpretation of transmission and reflection coefficients is lost away from the unitary slice.
Rather, $\mathcal{T}$ is the analytic continuation of a collider observable: it is the normalized coefficient with which the transmitted excitation contributes to the stress-tensor flux detector.

There is nevertheless a useful microscopic interpretation when the complex defect CFT is realized by a non-Hermitian Hamiltonian.
In a biorthogonal formulation, local stress-tensor or Hamiltonian-density expectation values can themselves be complex.
The complex coefficient $\mathcal{T}$ can then be measured as a ratio of complex-valued biorthogonal energy fluxes, as demonstrated for the complex Dirac interface in Sec.~\ref{sec:lattice}.
Thus ``complex energy flux'' always means a complex expectation value of an operator in the biorthogonal state, not a positive energy observable in a unitary Hilbert space.
This distinction is important: the CFT object is a transmission coefficient, while the lattice protocol provides one operational way of measuring the same analytically continued response.

Although positivity is lost, the conformal gluing condition still enforces the complexified flux-balance identity
\begin{equation}
\mathcal{R}(\phi_0)+\mathcal{T}(\phi_0)=1.
\label{eq:complex_flux_balance}
\end{equation}
Equivalently,
\begin{equation}
\mathrm{Re}\,\mathcal{R}+\mathrm{Re}\,\mathcal{T}=1,
\qquad
\mathrm{Im}\,\mathcal{R}+\mathrm{Im}\,\mathcal{T}=0.
\label{eq:complex_flux_balance_components}
\end{equation}
These are the real and imaginary parts of the complexified flux-balance relation following from the conformal gluing condition.
When represented by biorthogonal energy-flux measurements, they express algebraic conservation of the full complex flux.

Two nonunitary real-valued loci are especially instructive.
Along the vertical line through a topological point,
\begin{equation}
\phi_0=\frac{\pi}{4}+i\eta,
\end{equation}
one finds
\begin{equation}
\mathcal{T}=\cosh^2(2\eta),
\qquad
\mathcal{R}=-\sinh^2(2\eta).
\label{eq:topological_vertical_TR}
\end{equation}
Thus the transmission coefficient exceeds the unitary value, while the reflection coefficient carries a negative weight.
Conversely, along the vertical line through a factorizing point,
\begin{equation}
\phi_0=i\eta,
\end{equation}
one has
\begin{equation}
\mathcal{R}=\cosh^2(2\eta),
\qquad
\mathcal{T}=-\sinh^2(2\eta).
\label{eq:factorizing_vertical_TR}
\end{equation}

%%%%%%%%%%%%%%%%%%%%%%%%%%%%%%%%%%%%%%%%%%%%%%%%%%%%%%%%%%%%%%%%%%%%%%%%%%%%%%%%
\subsection{Entanglement and effective central charge}
\label{sec:IsingDefects:ceff}
%%%%%%%%%%%%%%%%%%%%%%%%%%%%%%%%%%%%%%%%%%%%%%%%%%%%%%%%%%%%%%%%%%%%%%%%%%%%%%%%

A second observable that can be continued and tested numerically is the entanglement entropy across the interface.
For a real Ising interface, the replica computation gives a logarithmic scaling of the form \cite{Brehm:2015lja},
\begin{equation}
S=\frac{c_{\rm eff}(s)}{6}\log L+O(1),
\qquad
s=|\sin(2\phi_0)|=\sqrt{\mathcal{T}}\in[0,1].
\label{eq:EE_sigma_real}
\end{equation}
With this convention $c_{\rm eff}=0$ for a factorizing defect and $c_{\rm eff}=c_{\rm Ising}=1/2$ for a topological defect.
The exact result can be written as
\begin{align}
c_{\rm eff}(s)
&=\frac{s}{2}-\frac{1}{2}
-
\frac{3}{\pi^2}
\Big[
(s+1)\log(s+1)\log s
+(s-1)\operatorname{Li}_2(1-s)
+(s+1)\operatorname{Li}_2(-s)
\Big],
\label{eq:sigma_dilog}
\end{align}
where $\operatorname{Li}_2(s)$ is the dilogarithm function and the branch is chosen so that the expression is real and positive for $0<s<1$.

The complex defect manifold makes the analytic structure of this quantity explicit.
Since the real formula depends on $|\sin(2\phi_0)|$, its holomorphic continuation requires a choice of a local square root of the transmission coefficient,
\begin{equation}
s^2=\mathcal{T}(\phi_0)=\sin^2(2\phi_0).
\label{eq:s_square_T}
\end{equation}
In a patch containing the identity topological defect at $\phi_0=\pi/4$, we take $s=\sin(2\phi_0)$.
A patch around the other topological point $\phi_0=3\pi/4$ requires the opposite sign if one wants $s=+1$ there.
Thus the absolute value in the real theory is replaced, after complexification, by a sheet choice.
The principal sheet is obtained by starting from the unitary interval $0<s<1$, imposing
\begin{equation}
c_{\rm eff}(0)=0,
\qquad
c_{\rm eff}(1)=\frac12,
\label{eq:ceff_endpoints}
\end{equation}
and analytically continuing without crossing a branch cut.

Several local features follow directly from Eq.~\eqref{eq:sigma_dilog}.
Near a factorizing point $s=0$, the analytically continued entanglement coefficient has the nonanalytic expansion
\begin{equation}
c_{\rm eff}(s)
=
\frac{3}{\pi^2}s^2
\left(\log\frac{1}{s}+\frac32\right)
+O\!\left(s^4\log s\right).
\label{eq:ceff_small_s}
\end{equation}
Thus the factorizing defect is not merely a zero of $c_{\rm eff}$ but a logarithmic branch point of the entanglement observable.
Analytic continuation around this point changes the branch as
\begin{equation}
\log s\longrightarrow \log s+2\pi i\ell,
\qquad \ell\in\mathbb{Z},
\end{equation}
and hence
\begin{equation}
c_{\rm eff}^{(\ell)}(s)
=
\frac{3}{\pi^2}s^2
\left(\log\frac{1}{s}+\frac32-2\pi i\ell\right)
+O\!\left(s^4\log s\right).
\label{eq:ceff_small_s_sheet}
\end{equation}
A counterclockwise winding of $s$ around the factorizing point induces the monodromy
\begin{equation}
c_{\rm eff}^{(\ell+1)}(s)-c_{\rm eff}^{(\ell)}(s)
=
-\frac{6i}{\pi}s^2+O(s^4).
\label{eq:ceff_monodromy}
\end{equation}
For $s=re^{i\theta}$ on a fixed sheet, the leading algebraic factor $s^2$ approximately doubles the local phase, $c_{\rm eff}\propto e^{2i\theta}$, while the logarithm records the winding history around the branch point.
The different branches therefore correspond to different analytic continuations of the replica free energy on the complex defect conformal manifold.

By contrast, the topological point $s=1$ is regular on the principal sheet:
\begin{equation}
c_{\rm eff}(s)
=\frac12+\frac34(s-1)+O\!\left((s-1)^2\right).
\label{eq:ceff_near_topological}
\end{equation}
Along the nonunitary real locus $\phi_0=\pi/4+i\eta$, where $s=\cosh(2\eta)>1$, the effective central charge remains real but immediately exceeds the unitary value $c_{\rm Ising}=1/2$.
At large positive $s$ on the same sheet,
\begin{equation}
c_{\rm eff}(s)=\frac32 s+o(s),
\qquad s\to +\infty.
\label{eq:ceff_large_s}
\end{equation}

Using the stress-tensor normalization in Eq.~\eqref{eq:cLR_transmission_Ising}, the same sheet choice gives
\begin{equation}
c_{LR}=\frac12s^2.
\label{eq:cLR_Ising}
\end{equation}
For real unitary defects this is compatible with the proposed bound $0\le c_{LR}\le c_{\rm eff}\le c$ \cite{Karch2024,Karch2023}.
Complexification removes the positivity assumptions behind this inequality.
Indeed, along $s=\cosh(2\eta)>1$ both $c_{LR}$ and $c_{\rm eff}$ are real, but $c_{LR}\sim s^2/2$ grows parametrically faster than $c_{\rm eff}\sim3s/2$.
Thus the bound does not extend even to this real nonunitary locus, where all quantities remain real on the chosen sheet.

%%%%%%%%%%%%%%%%%%%%%%%%%%%%%%%%%%%%%%%%%%%%%%%%%%%%%%%%%%%%%%%%%%%%%%%%%%%%%%%%
\subsection{Distinguished regions on the complex defect manifold}
\label{sec:IsingDefects:regions}
%%%%%%%%%%%%%%%%%%%%%%%%%%%%%%%%%%%%%%%%%%%%%%%%%%%%%%%%%%%%%%%%%%%%%%%%%%%%%%%%

\begin{figure}[t]
 \begin{center}
  \includegraphics[width=12.0cm,clip]{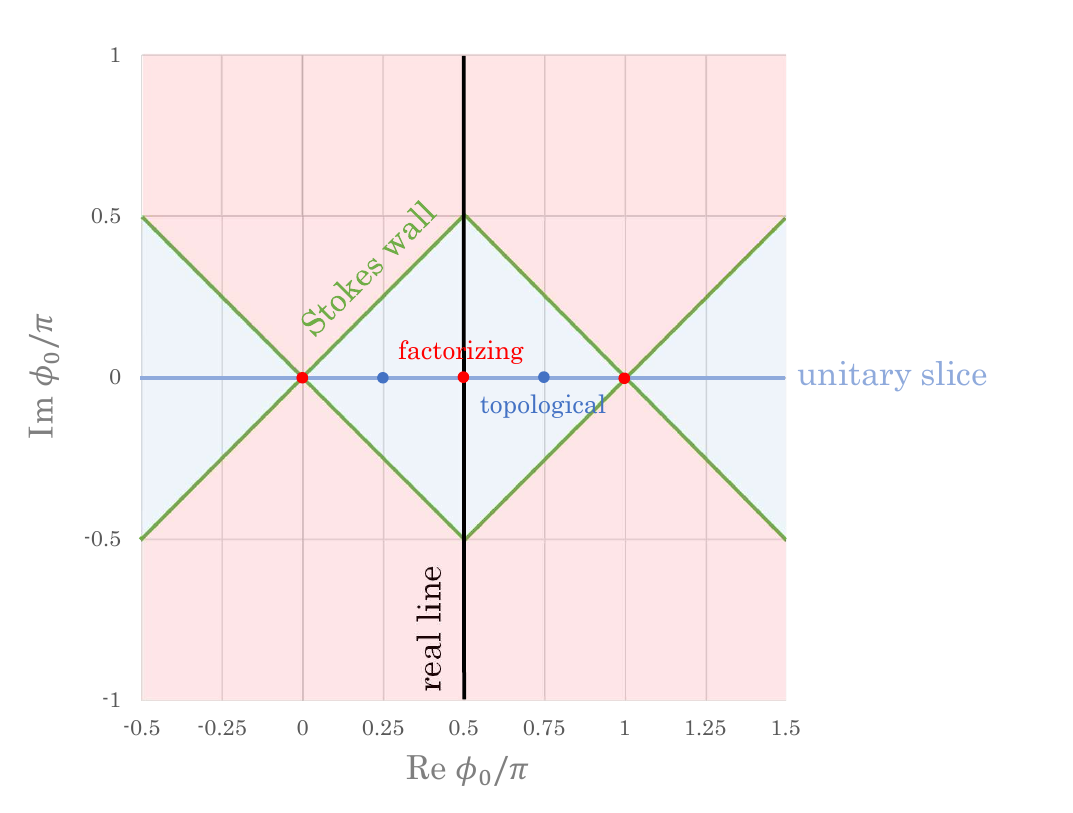}
 \end{center}
\caption{
Complex defect conformal manifold in the complex $\phi_0$-plane.
The real axis $\mathrm{Im}\,\phi_0=0$ is the unitary slice of the defect family.
Writing $\phi_0/\pi=x+iy$ and $\delta(x)=\min_{m\in\mathbb{Z}}|x-m|$, the blue region $|y|<\delta(x)$ is the decaying domain for the tower of the defect operators, bounded by the Stokes walls $|y|=\delta(x)$ (the green lines), which separate it from the amplifying domain $|y|>\delta(x)$ (the red region).
Vertical lines at $\mathrm{Re}\,\phi_0=m\pi/4$ indicate nonunitary loci where the transmission coefficient $\mathcal{T}(\phi_0)$ is real.
Special points on the unitary slice highlight the topological points $\phi_0=\pi/4,\,3\pi/4$ (totally transmitting) and the factorizing points $\phi_0=0,\,\pi/2,\,\pi$ (totally reflecting).
}
 \label{fig:CDCM}
\end{figure}

The analytic formulas above allow us to identify several distinguished regions and loci on the complex defect conformal manifold.
They are summarized in Fig.~\ref{fig:CDCM}.

\subsubsection{Unitary slice and nonunitary domain}
\label{sec:IsingDefects:unitary_slice}

The original unitary defect manifold is the real line
\begin{equation}
\mathrm{Im}\,\phi_0=0.
\label{eq:unitary_slice_defect}
\end{equation}
On this slice, $\mathcal{R},\mathcal{T}\in[0,1]$, and the ANEC collider observable of Sec.~\ref{sec:IsingDefects:energytransmission} reduces to an ordinary energy transmission coefficient.
Away from this slice, $\mathcal{R}$ and $\mathcal{T}$ are instead complex stress-tensor coefficients.
When a non-Hermitian lattice realization exists, they may be measured as ratios of complex-valued biorthogonal fluxes, but not as probabilities.

\subsubsection{Spectral growth/decay region and a ``Stokes wall''}
\label{sec:IsingDefects:ReDelta_wall}

Another sharp distinction concerns whether the smallest real part in the tower \eqref{eq:defect_spectrum_moving} is positive.
The protected identity sector is present independently of this tower.
Define $\phi_0/\pi=x+iy$ and the distance to the nearest integer
\begin{equation}
\delta(x):=\min_{m\in\mathbb{Z}}|x-m|\in\left[0,\frac12\right].
\label{eq:delta_x}
\end{equation}
Then
\begin{equation}
\mathrm{Re}\,\Delta_{\min}(x,y)=2\big(\delta(x)^2-y^2\big).
\label{eq:ReDelta_min}
\end{equation}
This leads to three regions:
\begin{align}
|y|<\delta(x)
&\quad\Rightarrow\quad
\mathrm{Re}\,\Delta_{\min}>0
\qquad \text{(decaying)},
\label{eq:region_decay}
\\
|y|=\delta(x)
&\quad\Rightarrow\quad
\mathrm{Re}\,\Delta_{\min}=0
\qquad \text{(purely oscillatory)},
\label{eq:region_boundary}
\\
|y|>\delta(x)
&\quad\Rightarrow\quad
\mathrm{Re}\,\Delta_{\min}<0
\qquad \text{(amplifying)}.
\label{eq:region_amplify}
\end{align}
The wall $|y|=\delta(x)$ is a natural Stokes wall on the complex manifold: it is where the asymptotic dominance structure of correlators in this tower changes qualitatively.

\subsubsection{Real loci of defect data}
\label{sec:IsingDefects:real_loci}

As in the complex Gaussian example, symmetry and analytic continuation can enforce the reality of certain observables even away from the unitary slice.
For the Ising defect, Eq.~\eqref{eq:complex_T_explicit} gives
\begin{equation}
\mathrm{Im}\,\mathcal{T}(u+i\eta)
=
\frac12\sin(4u)\sinh(4\eta).
\label{eq:Im_T_real_loci}
\end{equation}
Hence $\mathcal{T}$ is real either on the unitary slice $\eta=0$ or on the vertical lines
\begin{equation}
\mathrm{Re}\,\phi_0=u=\frac{m\pi}{4},
\qquad m\in\mathbb{Z}.
\label{eq:real_loci_T_all}
\end{equation}
These nonunitary real loci contain the vertical continuations of the topological and factorizing points described in Eqs.~\eqref{eq:topological_vertical_TR} and \eqref{eq:factorizing_vertical_TR}.
On them the collider response is real, but it need not lie in the unitary interval $[0,1]$.

One of these loci also has a simple symmetry explanation.
Complex conjugation gives
\begin{equation}
\mathcal{T}(\phi_0)^\ast=\mathcal{T}(\phi_0^\ast).
\label{eq:RT_conj}
\end{equation}
Fusion with the topological $\varepsilon$-defect implements $\phi_0\mapsto\pi-\phi_0$ on the defect family and leaves $\mathcal{T}$ invariant.
Therefore, on the fixed line
\begin{equation}
\phi_0^\ast=\pi-\phi_0
\qquad\Longleftrightarrow\qquad
\mathrm{Re}\,\phi_0=\frac{\pi}{2},
\label{eq:real_locus_defect}
\end{equation}
one has $\mathcal{T}(\phi_0)^\ast=\mathcal{T}(\pi-\phi_0)=\mathcal{T}(\phi_0)$.

\subsubsection{Topological points and factorizing cusps}
\label{sec:IsingDefects:special_points}

The points in Eqs.~\eqref{eq:lambda_topological_identity}, \eqref{eq:lambda_topological_second}, and \eqref{eq:lambda_factorizing} remain distinguished after complexification, although only the factorizing points become branch points of the entanglement observable.
They are characterized by algebraic properties that survive analytic continuation.
\begin{itemize}
\item \textbf{Topological points ($\mathcal{R}=0$):} the defect is totally transmitting and can be moved without affecting correlators away from its endpoints.  In the folded picture, this corresponds to a maximally symmetric boundary condition.
\item \textbf{Factorizing points ($\mathcal{T}=0$):} the defect is totally reflecting and effectively splits space into two decoupled halves.  These are also the logarithmic branch points of $c_{\rm eff}$ in the complexified entanglement observable.
\end{itemize}

%%%%%%%%%%%%%%%%%%%%%%%%%%%%%%%%%%%%%%%%%%%%%%%%%%%%%%%%%%%%%%%%%%%%%%%%%%%%%%%%%%%%%%%%%%%%%%
%%%%%%%%%%%%%%%%%%%%%%%%%%%%%%%%%%%%%%%%%%%%%%%%%%%%%%%%%%%%%%%%%%%%%%%%%%%%%%%%%%%%%%%%%%%%%%
\section{Lattice Realization of the Complex Defect Conformal Manifold}
\label{sec:lattice}
%%%%%%%%%%%%%%%%%%%%%%%%%%%%%%%%%%%%%%%%%%%%%%%%%%%%%%%%%%%%%%%%%%%%%%%%%%%%%%%%%%%%%%%%%%%%%%
%%%%%%%%%%%%%%%%%%%%%%%%%%%%%%%%%%%%%%%%%%%%%%%%%%%%%%%%%%%%%%%%%%%%%%%%%%%%%%%%%%%%%%%%%%%%%%

To provide numerical verification for the complex defect conformal manifold described above, we study a lattice realization of the Ising CFT with a complexified defect. Specifically, we consider the transverse-field Ising chain with an energy defect \cite{Oshikawa1996,Roy:2021xul}, governed by the Hamiltonian 
\begin{equation} 
H = -\frac{1}{2} \sum_{j=1}^{L-1} \sigma_j^z \sigma_{j+1}^z - \frac{1}{2} \sum_{j=1}^L \sigma_j^x + \frac{1-b_\epsilon}{2} \sigma_{i_0}^z \sigma_{i_0+1}^z, 
\label{eq:Lat_Ham} 
\end{equation} 
where $L$ is the total number of sites, $i_0=L/2$ denotes the central bond containing the defect, and open boundary conditions are imposed. In previous studies the defect coupling $b_\epsilon$ was taken real in $[0,1]$, whereas here we extend $b_\epsilon$ to complex values, rendering $H$ non-Hermitian.

\subsection{Biorthogonal correlation matrix and entanglement}

Under the Jordan--Wigner transformation, the Hamiltonian \eqref{eq:Lat_Ham} maps to a free Majorana fermion chain, where all physical information can be encoded in the two-point correlation matrix. In the non-Hermitian setting, we use the biorthogonal basis of left and right eigenstates, $\langle \Psi_L|$ and $|\Psi_R\rangle$, to define correlation functions. The two-point Majorana correlation matrix is defined by $ C_{ij} = \frac{\langle \Psi_L | \gamma_i \gamma_j | \Psi_R \rangle}{\langle \Psi_L | \Psi_R \rangle}$, where $\gamma_i$ are the Majorana operators \cite{Peschel:2002yqj,Eisler:2009vye}. Upon partitioning the chain into $A \cup\bar{A}$, the reduced density matrix obtained from the biorthogonal eigenstates $\rho_A=\text{Tr}_{\bar{A}}|\Psi_R \rangle\langle \Psi_L|$ is similarly non-Hermitian, and its eigenvalues $\rho_k$ are generally complex, leading to negative or complex scaling behavior \cite{Chang:2019jcj,Shimizu:2025kse}. This construction is also closely related to the definition of pseudo-entropy within high energy contexts \cite{Nakata:2020luh,Mollabashi:2020yie,Doi:2022iyj}. Consequently, the von Neumann entanglement entropy $ S=-\sum_k \rho_k \ln \rho_k$ and expectation values of local operators become complex-valued quantities.

\subsection{Bulk-defect correlator and defect scaling dimensions}

To extract the defect scaling dimension in this lattice model, we compute the biorthogonal bulk-defect correlator in the above open chain. We define $ G(l) = \frac{\langle \Psi_L | \sigma^z_{i_0+1} \sigma^z_{i_0+l} | \Psi_R \rangle}{\langle \Psi_L | \Psi_R \rangle},$  where one $\sigma^z$ operator is fixed at the site adjacent to the defect ($i_0+1$) and the other is a distance $l$ away in the bulk. Conformal invariance on the finite open chain implies that this correlator decays as a power law \cite{DiFrancesco:1997nk}: 
\begin{equation} 
G(l) \propto \frac{1}{\ell_{\rm eff}^{\Delta^{\sigma}_{\rm bulk} + \Delta^{\sigma}_{\rm defect}}}, \label{eq:scaling_law} 
\end{equation} 
where $\Delta^{\sigma}_{\rm bulk} = 1/8$ is the bulk scaling dimension of the Ising spin operator, and $\Delta^{\sigma}_{\rm defect}$ is the defect scaling dimension we aim to extract. The effective distance in the open chain of length $L$ is defined as $\ell_{\rm eff} = \tfrac{2L}{\pi}\sin\bigl(\frac{\pi l}{2L}\bigr)$.

\begin{figure}[t] 
\centering \includegraphics[width=12.0cm,clip]{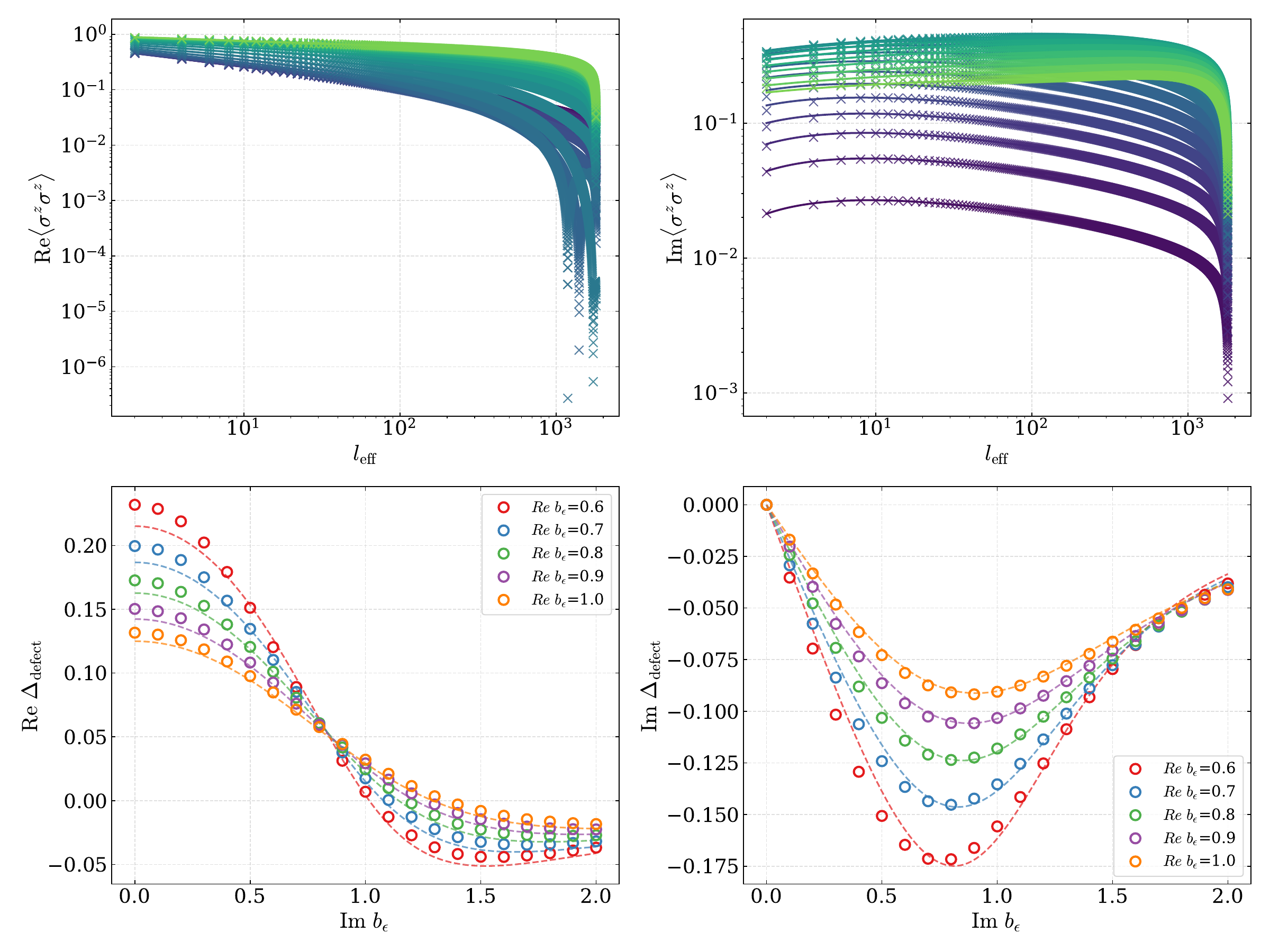} 
\caption{ Numerical analysis of the bulk-defect correlator based on biorthogonal eigenstates and the extracted defect scaling dimension $\Delta^{\sigma}_{\rm defect}$. \textbf{Top:} The real (left) and imaginary (right) parts of the correlator $G(l)=\langle\sigma^z_{i_0+1}\sigma^z_{i_0+l}\rangle$ versus the effective distance $\ell_{\rm eff}$, for fixed $\mathrm{Re}(b_\epsilon)=0.6$ and various $\mathrm{Im}(b_\epsilon)$ (distinguished by colors). Symbols are numerical data and solid lines are fits to the complex power law \eqref{eq:scaling_law}. 
\textbf{Bottom:} Real (left) and imaginary (right) parts of the extracted $\Delta^{\sigma}_{\rm defect}$ as functions of $\mathrm{Im}(b_\epsilon)$ (for several $\mathrm{Re}(b_\epsilon)$). Numerical results (circles) are compared to the analytic predictions (dashed lines). } \label{fig:defect_scaling} 
\end{figure}

The numerical results are presented in Fig.~\ref{fig:defect_scaling}. The top panels show the real and imaginary parts of $G(l)$ versus $\ell_{\rm eff}$ for a fixed $\mathrm{Re}(b_\epsilon)=0.6$ and various values of $\mathrm{Im}(b_\epsilon)$. The symbols are numerical data and the solid lines are fits to the complex power law \eqref{eq:scaling_law}. From the fits, we extract the complex defect dimension $\Delta^{\sigma}_{\rm defect}$. The bottom panels of Fig.~\ref{fig:defect_scaling} plot the real (left) and imaginary (right) parts of the extracted $\Delta^{\sigma}_{\rm defect}$ as functions of $\mathrm{Im}(b_\epsilon)$ for several $\mathrm{Re}(b_\epsilon)$. The numerical results (circles) closely follow the analytic continuation of the theoretical predictions (dashed lines) Eq.~\eqref{eq:Delta_sigma_defect_b}. We find excellent agreement in both components across the complex defect manifold. This confirms that the complex coupling $b_\epsilon$ acts as a marginal deformation, continuously tuning the defect scaling dimension along the manifold.

\subsection{Scaling of complex entanglement entropy and effective central charge}

We next examine the entanglement entropy of a subsystem when the entanglement cut passes through the defect. Defect CFT predicts that at criticality, the entropy obeys a universal logarithmic scaling \cite{Calabrese:2009qy}. In particular, for a half-chain (subsystem length $L/2$) the entropy should follow 
\begin{equation} 
S(L/2) = \frac{c_{\rm eff}(b_\epsilon)}{6} \ln L + \text{const}, \label{eq:EE_scaling} 
\end{equation} 
where $c_{\rm eff}$ is an effective central charge characterizing the transmission of entanglement across the defect \cite{Cardy:1987vr,Eisler:2012xry,Wen:2017smb,Roy:2021xul,Rogerson:2022yim,Karch2023}. For the Ising defect with real coupling, an analytic formula for $c_{\rm eff}$ was derived in \cite{Eisler:2010vwa,Brehm:2015lja}. This formula can be naturally extended to complex $b_\epsilon$ by analytic continuation. Explicitly, one can write \begin{equation} c_{\rm eff}(b_\epsilon) = \frac{1}{2}t - \frac{1}{2} - \frac{3}{\pi^2}\Bigl[(t+1)\ln(t+1)\ln t+(t-1)\mathrm{Li}_2(1-t)+(t+1)\mathrm{Li}_2(-t)\Bigr], \label{eq:ceff_theory} \end{equation} where the parameter $ t = \sin\bigl[2\arctan(1/b_\epsilon)\bigr] $ is related to the defect strength. In the unitary case $b_\epsilon\in[0,1]$, $c_{\rm eff}$ varies from 0 to $1/2$ as the defect interpolates from reflective to transparent. Generically, the expression \eqref{eq:ceff_theory} defines a complex-valued $c_{\rm eff}$ for complex $b_\epsilon$.

\begin{figure}[t] 
\centering \includegraphics[width=12.0cm,clip]{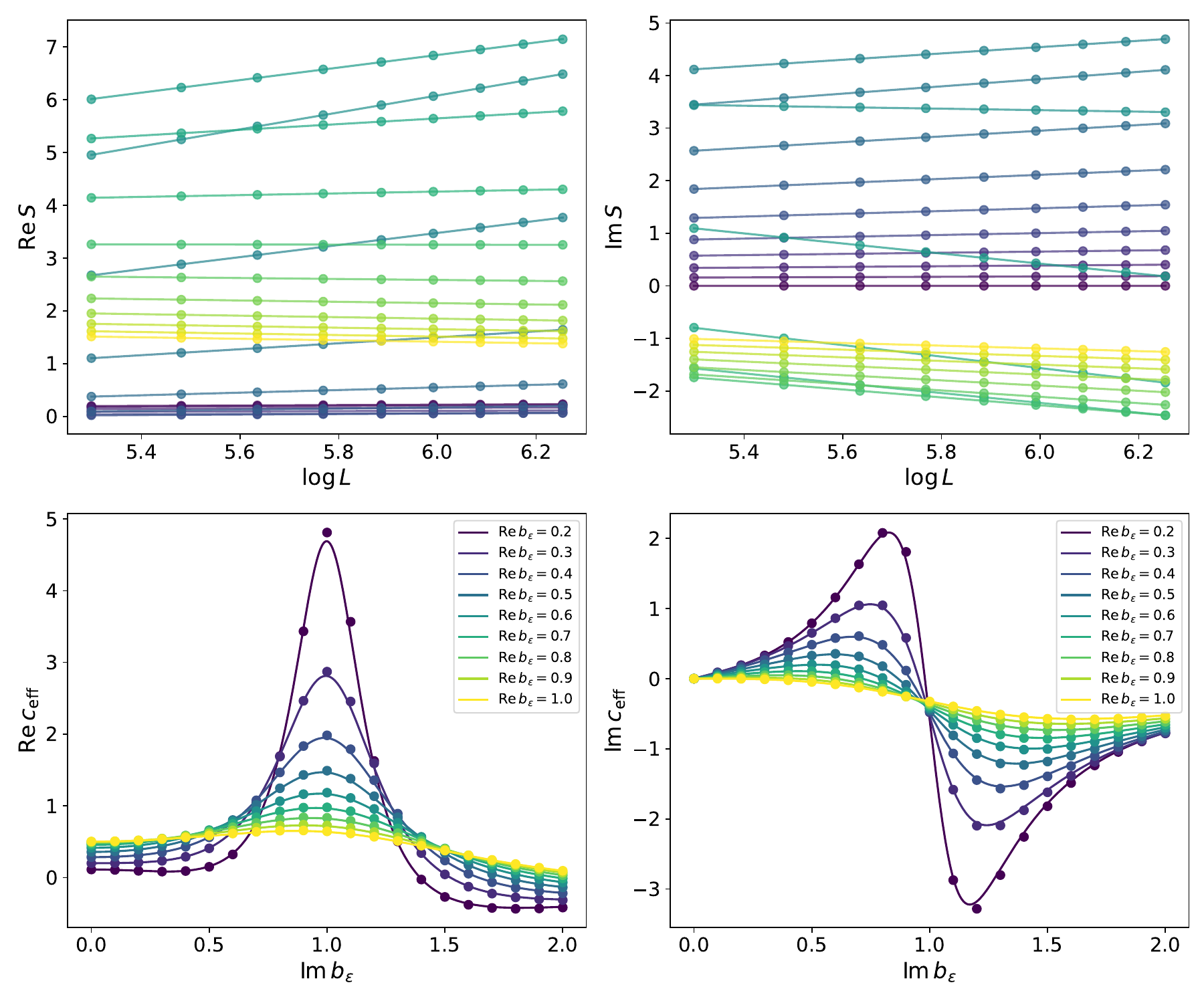} \caption{ Numerical analysis of the entanglement entropy scaling and the effective central charge $c_{\rm eff}$. \textbf{Top row:} The real (left) and imaginary (right) parts of the entanglement entropy at the defect ($l=L/2$) as functions of $\ln L$, for several values of $\mathrm{Im}(b_\epsilon)$ at fixed $\mathrm{Re}(b_\epsilon)=0.5$. Solid lines are linear fits $S\sim (c_{\rm eff}/6)\ln L + a$, confirming the logarithmic law. \textbf{Bottom row:} Comparison of the real (left) and imaginary (right) parts of the extracted $c_{\rm eff}$ (symbols) with the analytic continuation of Eq.~\eqref{eq:ceff_theory} (solid curves) across the complex defect manifold. Excellent agreement is seen in both components. } 
\label{fig:ceff_scaling} 
\end{figure}

Figure~\ref{fig:ceff_scaling} shows the results of a finite-size scaling analysis up to $L=520$. In the top panels, both $\mathrm{Re}[S(L/2)]$ and $\mathrm{Im}[S(L/2)]$ grow linearly with $\ln L$ for various complex defect strengths, confirming the logarithmic scaling form \eqref{eq:EE_scaling}. The bottom panels compare the real and imaginary parts of the extracted $c_{\rm eff}$ (symbols) to the analytic continuation of Eq.~\eqref{eq:ceff_theory} (curves). The agreement is excellent across the complex manifold, demonstrating that the logarithmic entanglement scaling persists even in the non-Hermitian regime. These findings confirm that conformal symmetry is preserved and that the complex defect coupling also acts as a marginal perturbation tuning the effective central charge throughout the defect manifold.

\begin{figure}[t] 
\centering \includegraphics[width=12.0cm,clip]{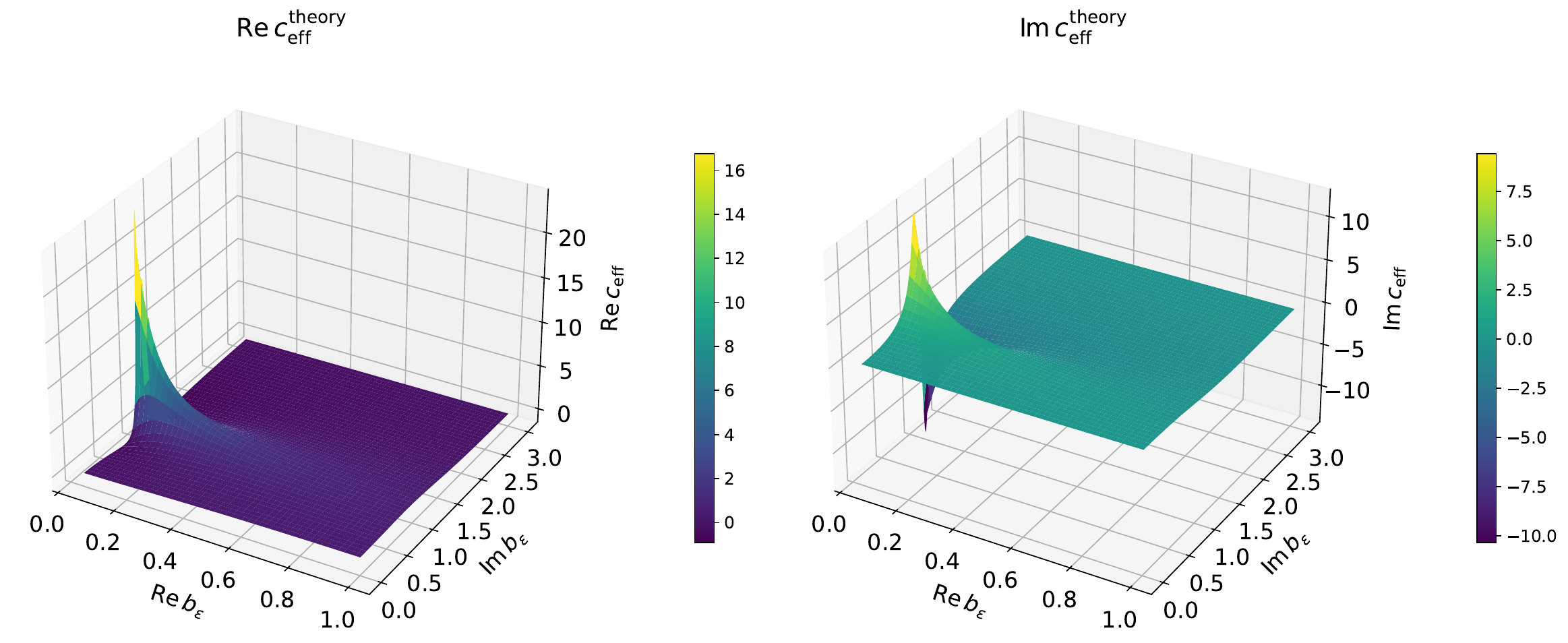} \caption{ Theoretical landscape of the complex effective central charge $c_{\rm eff}$ over the complex coupling plane $b_\epsilon$. The left and right panels show the real and imaginary parts of $c_{\rm eff}$, respectively, as given by the analytic continuation of Eq.~\eqref{eq:ceff_theory}. } \label{fig:ceff_3d_landscape} 
\end{figure}

Figure~\ref{fig:ceff_3d_landscape} visualizes $c_{\rm eff}(b_\epsilon)$ as a continuous surface over the complex coupling plane. The smooth variation indicates a well-defined “landscape” of defect critical behavior in the region shown. Together with the correlator data, these results support the conclusion that this lattice model \eqref{eq:Lat_Ham} realizes the predicted complex defect conformal manifold: both the defect scaling dimension and the effective central charge match the analytic continuation of the defect CFT predictions.

\subsection{Transmission coefficient on a complex Dirac fermion interface}

Finally, we study a different model which also hosts a continuous defect manifold: a critical free Dirac (tight-binding) chain with a conformal interface \cite{Eisler:2010vwa,Wen:2017smb,Barad:2025gba}. We take two identical chains of length $N$, joined at the center (total length $L=2N$) with an interface at the middle bond. The Hamiltonian can be written as
\begin{equation}
H=\frac{1}{2}\sum_{m,n=1}^{2N} H_{mn}\, c_m^\dagger c_n,
\label{eq:dirac_interface_hamiltonian}
\end{equation}
where $c_m^\dagger$ and $c_n$ are the fermionic creation and annihilation operators, and the nonvanishing matrix elements are given by
\begin{equation}
H_{m,m+1}=H_{m+1,m}=
\begin{cases}
-1, & m\neq N,\\
-\lambda, & m=N,
\end{cases}
\qquad
H_{N,N}=-H_{N+1,N+1}=\sqrt{1-\lambda^2}.
\label{eq:interface_matrix_elements}
\end{equation} Here $\lambda$ parametrizes the defect: for real $\lambda\in[0,1]$, the interface is unitary, interpolating between a fully reflecting ($\lambda=0$) and a fully transmitting ($\lambda=1$) defect. In that unitary regime the transmission coefficient is $\mathcal{T}=\lambda^2$ \cite{Quella2006,Eisler:2010vwa}. We now analytically continue $\lambda$ to complex values, making $H$ non-Hermitian. As before, we work in the biorthogonal basis: we prepare the initial state $\rho_{\rm LR}(0)=|R_0\rangle\langle L_0|/\langle L_0|R_0\rangle$ from the ground states of $H$, and evolve it under biorthogonal time evolution governed by $\rho_{\rm LR}(t)=e^{-iHt} \rho_{\rm LR}(0) e^{iHt}$. Expectation values of operators are then computed as $\langle \mathcal{O}\rangle_{\rm LR}=\langle L(t)|\mathcal{O}|R(t)\rangle/\langle L(t)|R(t)\rangle$.

To measure the transmission coefficient on the defect manifold, we employ a local joining quench protocol studied in \cite{Barad:2025gba}. Initially, at $t<0$, the two parts of the chain are disconnected by setting $H_{l_0,l_0+1}=H_{l_0+1,l_0}=0$. At $t=0$, they are joined at site $l_0$ by turning on the coupling at the junction. This creates a localized energy pulse, which then propagates ballistically away from the joining point. After a time of order $t\sim d/v$, where $d$ is the distance between the joining site and the defect and $v$ is the characteristic velocity, the right-moving pulse reaches the defect at $l=L/2$.

In the unitary regime, the incoming pulse is split into transmitted and reflected parts. On the complex defect manifold, the same scattering picture continues to hold, but the transmitted and reflected energy densities are generally complex. We therefore define the transmission and reflection coefficients from the energy flux as
\begin{equation}
\mathcal{T}_{\rm num}
=\frac{E_{\rm trans}}{E_{\rm inc}},
\qquad
\mathcal{R}_{\rm num}
=\frac{E_{\rm refl}}{E_{\rm inc}},
\end{equation}
where $E_{\rm inc}$, $E_{\rm trans}$, and $E_{\rm refl}$ are obtained by integrating the local energy density over spacetime windows associated with the incoming, transmitted, and reflected wave packets, respectively \cite{Quella2006,Meineri2019}. In the complex case, one may evaluate these quantities separately for the real and imaginary parts of the energy density, or equivalently as the full complex flux. We expect $\mathcal{T}_{\rm num}$ to follow the analytic continuation of $
\mathcal{T}^{\rm (complex)}=\lambda^2,$
which provides a direct physical interpretation of the complex transmission coefficient.

\begin{figure}[t] 
\centering \includegraphics[width=12.0cm,clip]{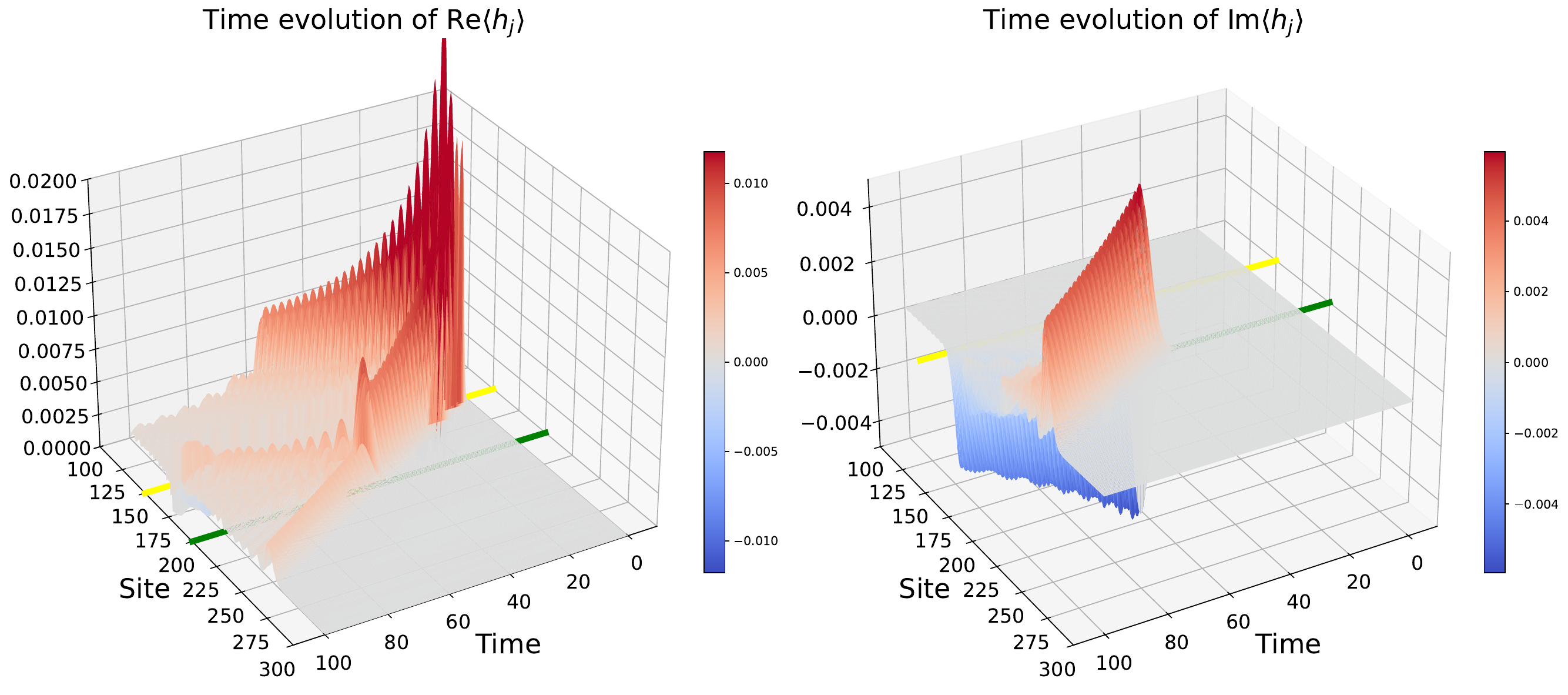} \caption{ Spacetime evolution of the local energy density after a local quench in the presence of a complex defect. The $x-$axis is lattice site and the $y-$axis is time. The left panel shows $\mathrm{Re}\langle h_j\rangle_{\rm LR}$ and the right panel shows $\mathrm{Im}\langle h_j\rangle_{\rm LR}$. The yellow line marks the quench (joining) site and the green line marks the location of the defect. After the joining quench at $t=0$, an energy pulse is generated and propagates away from the quench. When the pulse reaches the defect, it splits into transmitted and reflected components. For the non-Hermitian defect, an imaginary part of the energy density is simultaneously generated at the interface and propagates with both the transmitted and reflected pulses. } \label{fig:energy_evolution_complex_defect}
\end{figure}

For a representative choice such as $\lambda=0.9+0.5i$, Fig.~\ref{fig:energy_evolution_complex_defect} displays a typical spacetime evolution of the local energy density obtained from the joining quench protocol. The defect is located at $l=L/2=200$, while the quench is performed at $l_0=150$. The left panel shows $\mathrm{Re}\,\langle h_j\rangle_{\rm LR}(t)$ and the right panel shows $\mathrm{Im}\,\langle h_j\rangle_{\rm LR}(t)$. The yellow horizontal line marks the joining site, and the green horizontal line marks the defect position. After the quench at $t=0$, an energy pulse is created near the junction and propagates ballistically in both directions. Since the model is critical, the resulting wavefronts spread linearly in time and form the familiar light-cone pattern \cite{Calabrese:2005in,Calabrese:2016xau}.

When the right-moving wavefront reaches the complex defect, part of the energy is transmitted across the interface and part of it is reflected back. This scattering process is clearly visible in the real part of the local energy density. In particular, the real component behaves qualitatively as in the unitary case: the incoming pulse splits at the defect, the transmitted piece continues to the right, and the reflected piece travels back toward the left \cite{Barad:2025gba}. Thus, even after complexification, the defect still acts as a conformal interface with well-defined transmission and reflection processes \cite{Meineri2019}.

The most distinctive feature of the complex defect manifold appears in the imaginary part of the energy density. Since the Hamiltonian is no longer Hermitian for $\lambda\in\mathbb{C}$, neither the reduced density matrix nor the local energy density is constrained to remain real by definition. As a result, the defect not only modifies the amplitude of the outgoing wave packets, but also generates a nontrivial imaginary-energy component. This imaginary contribution is produced dynamically during the scattering process and then propagates away from the defect together with the real partner. Moreover, the reflected imaginary pulse typically carries the opposite sign, so that the total imaginary energy flux created after collision is still zero, as required by the total energy conservation.

\begin{figure}[t] 
\centering \includegraphics[width=12.0cm,clip]{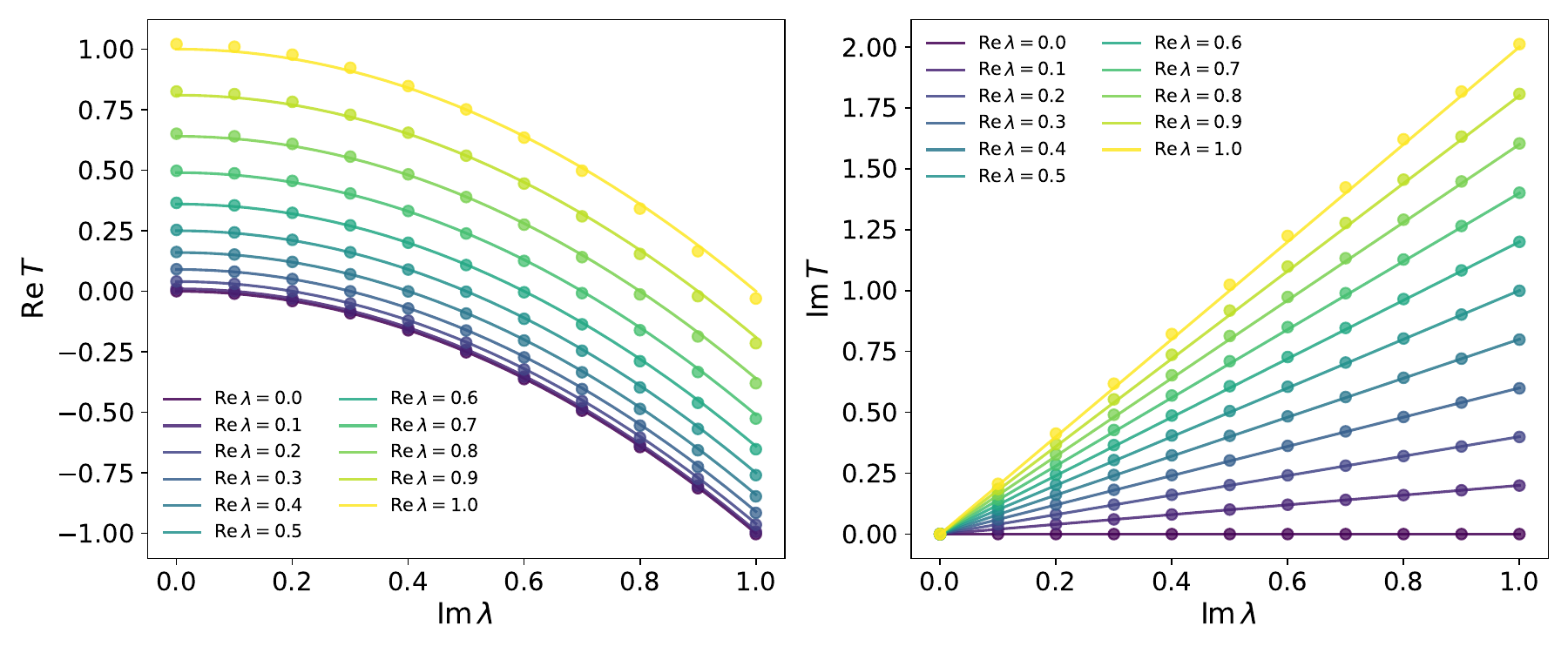} \caption{ Comparison of the numerical transmission coefficient with the analytic result on the complex defect manifold. The left and right panels show the real and imaginary parts of $\mathcal{T}$ as functions of $\mathrm{Im}(\lambda)$, for different $\mathrm{Re}(\lambda)$. The symbols are from the quench simulation and the solid curves are $\mathcal{T}=\lambda^2$. Excellent agreement is observed in both components across the entire parameter range. } 
\label{fig:transmission_coefficients_complex_defect} 
\end{figure}

To quantify this behavior, we compute $E_{\rm inc}$, $E_{\rm trans}$, and $E_{\rm refl}$ from the biorthogonal expectation values of the evolving state and form the ratio $\mathcal{T}_{\rm num}$. We find that the numerical result agrees very well with the analytically continued transmission coefficient $\mathcal{T}=\lambda^2$, even when $\lambda$ is complex. For the example $\lambda=0.9+0.5i$, this gives $
\mathcal{T}=\lambda^2=0.56+0.9i$,
which matches the numerical data $\mathcal{T}_{\rm num}(\lambda=0.9+0.5i)\simeq 0.5631+0.9186i$ to very good accuracy. This agreement indicates that the transmission coefficient remains a universal quantity on the complex defect conformal manifold, and that its physical meaning can still be read off from the scattering of a complex energy pulse. A comparison across the full defect manifold is shown in Fig.~\ref{fig:transmission_coefficients_complex_defect}.

An additional consistency check comes from the defect itself: the total complex energy carried by the transmitted and reflected wave packets reproduces the incoming energy, as expected from local conservation at the interface. Therefore, although the local energy density becomes complex and the notion of transmission must be generalized to the biorthogonal setting, the interface still admits a precise and physically meaningful characterization in terms of a complex transmission coefficient under nonunitary dynamics.

%%%%%%%%%%%%%%%%%%%%%%%%%%%%%%%%%%%%%%%%%%%%%%%%%%%%%%%%%%%%%%%%%%%%%%%%%%%%%%%%%%%%%%%%%%%%%%
%%%%%%%%%%%%%%%%%%%%%%%%%%%%%%%%%%%%%%%%%%%%%%%%%%%%%%%%%%%%%%%%%%%%%%%%%%%%%%%%%%%%%%%%%%%%%%
\section{Complex Boundary RG Flows and de Sitter Branes}
\label{sec:ComplexBoundaryRGFlows}
%%%%%%%%%%%%%%%%%%%%%%%%%%%%%%%%%%%%%%%%%%%%%%%%%%%%%%%%%%%%%%%%%%%%%%%%%%%%%%%%%%%%%%%%%%%%%%
%%%%%%%%%%%%%%%%%%%%%%%%%%%%%%%%%%%%%%%%%%%%%%%%%%%%%%%%%%%%%%%%%%%%%%%%%%%%%%%%%%%%%%%%%%%%%%

The examples discussed so far are complex conformal manifolds obtained by analytically continuing
exactly marginal couplings.  Since exactly marginal directions do not define RG trajectories, they
do not by themselves give rise to an ordering problem or to a monotonicity question.  RG flows
connected to complex fixed points are different: once the couplings are complexified, the positivity
assumptions underlying standard monotonicity theorems need not survive.

In this section we demonstrate this explicitly for the boundary $g$-theorem.  One might hope that,
although the Affleck--Ludwig theorem itself relies on unitarity, a weaker statement could persist
after complexification, for example as a monotonicity property of $\mathrm{Re}\,\log g$ or of
$|g|$.  We show that this is not the case in a simple AdS/BCFT setup
\cite{Takayanagi2011,Fujita2011,Kanda2023}.  A complex boundary RG flow can have a real
end-of-the-world brane embedding and a real value of $\log g$, yet its boundary entropy increases
from the UV to the IR.

The mechanism is the boundary counterpart of the holographic violation of the $c$-theorem in
complex CFTs \cite{Faedo2021}.  In the usual real flow, the scalar profile on the brane contributes
a positive term proportional to $(k^a\partial_a\varphi)^2$ to the null-energy condition.  After the
analytic continuation $\varphi=i\chi$, this term becomes
$-(k^a\partial_a\chi)^2$.  Thus the positivity input in the holographic proof is not merely lost
but sign-reversed, allowing the Affleck--Ludwig boundary entropy to increase along the complex
boundary RG flow.

A related construction appears in holographic PT-symmetric BCFTs, where an imaginary
brane-localized scalar is used to represent non-Hermitian but PT-symmetric boundary interactions
\cite{Maeda2026}.  There the PT symmetry provides an anti-linear reality structure which
keeps the holographic description controllable in the PT-unbroken regime.  Our setup is different:
we introduce a brane scalar potential and use it to describe relevant boundary RG flows.  We do not
impose a global PT symmetry.  Instead, we impose a weaker but explicit real-saddle condition, or
equivalently use a solution-generating map from ordinary real boundary RG flows.  This is the
precise sense in which the classical bulk calculation below is controlled.

%%%%%%%%%%%%%%%%%%%%%%%%%%%%%%%%%%%%%%%%%%%%%%%%%%%%%%%%%%%%%%%%%%%%%%%%%%%%%%%%%%%%%%%%%%%%%%
\subsection{AdS/BCFT with a brane-localized scalar}
\label{sec:ComplexBoundaryRGFlows:setup}
%%%%%%%%%%%%%%%%%%%%%%%%%%%%%%%%%%%%%%%%%%%%%%%%%%%%%%%%%%%%%%%%%%%%%%%%%%%%%%%%%%%%%%%%%%%%%%

We work in Lorentzian Poincar\'e AdS$_3$,
\begin{equation}
ds^2=
\frac{L^2}{z^2}
\left(
dz^2-dt^2+dx^2
\right),
\label{eq:AdS3_Poincare_complex_flow}
\end{equation}
and mostly set $L=1$ below.  The end-of-the-world brane $Q$ is described by
\begin{equation}
Q:\qquad x=X(z),
\label{eq:EOW_embedding_complex_flow}
\end{equation}
where increasing $z$ corresponds to moving from the UV to the IR of the boundary RG flow.

We consider an AdS/BCFT model with a scalar field localized on $Q$ \cite{Takayanagi2011,Fujita2011,Kanda2023},
\begin{equation}
I_Q
=
-\frac{1}{8\pi G_N}
\int_Q d^2\sigma \sqrt{-h}\,
\left[
K
-
V_Q(\varphi)
-
h^{ab}\partial_a\varphi\partial_b\varphi
\right].
\label{eq:brane_scalar_action_complex_flow}
\end{equation}
Here $Q$ is the two-dimensional end-of-the-world brane with intrinsic
coordinates \(\sigma^a\).  We denote by \(h_{ab}\) the induced metric on
$Q$, by \(h=\det h_{ab}\) its determinant, and by \(K=h^{ab}K_{ab}\)
the trace of the extrinsic curvature.  The scalar field \(\varphi\) is
localized on $Q$, \(V_Q(\varphi)\) is its brane potential, and
\(G_N\) is the three-dimensional Newton constant.
For a constant scalar configuration, the value of the potential acts as an effective brane tension,
\begin{equation}
T_{\rm eff}=V_Q(\varphi_\ast).
\label{eq:effective_tension_complex_flow}
\end{equation}
Thus critical points of $V_Q$ describe boundary fixed points, while interpolating profiles
$\varphi(z)$ describe boundary RG flows.

The Neumann condition on the brane is
\begin{equation}
K_{ab}-K h_{ab}
+
\left(
V_Q(\varphi)+h^{cd}\partial_c\varphi\partial_d\varphi
\right)h_{ab}
-
2\partial_a\varphi\partial_b\varphi
=0,
\label{eq:Neumann_complex_flow}
\end{equation}
and the equation of motion for the scalar field is
\begin{equation}
2\nabla_Q^2\varphi=V_Q'(\varphi).
\label{eq:scalar_EOM_complex_flow}
\end{equation}
At a subcritical fixed point,
\begin{equation}
|T_{\rm eff}|L<1,
\label{eq:subcritical_condition}
\end{equation}
the brane is an AdS$_2$ brane,
\begin{equation}
X(z)=z\sinh\rho_\ast,
\qquad
T_{\rm eff}L=\tanh\rho_\ast,
\label{eq:AdS2_fixed_point_brane}
\end{equation}
up to the choice of orientation of the retained bulk region.

%%%%%%%%%%%%%%%%%%%%%%%%%%%%%%%%%%%%%%%%%%%%%%%%%%%%%%%%%%%%%%%%%%%%%%%%%%%%%%%%%%%%%%%%%%%%%%
\subsection{Real-saddle condition}
\label{sec:ComplexBoundaryRGFlows:reality}
%%%%%%%%%%%%%%%%%%%%%%%%%%%%%%%%%%%%%%%%%%%%%%%%%%%%%%%%%%%%%%%%%%%%%%%%%%%%%%%%%%%%%%%%%%%%%%

A generic complex boundary source would make the brane stress tensor complex.  Then the Neumann
condition \eqref{eq:Neumann_complex_flow} would generally require a complex embedding $X(z)$ or a
complex bulk metric.  In such a situation, a real inequality for $\log g$ would not be meaningful
without specifying an additional prescription for complex saddles.

We therefore restrict to a real-saddle sector.  We take
\begin{equation}
\varphi=i\chi,
\qquad
\chi\in\mathbb{R},
\label{eq:imaginary_scalar_real_saddle}
\end{equation}
and choose the potential so that the imaginary axis is closed under the equations of motion:
\begin{equation}
V_Q(i\chi)\in\mathbb{R},
\qquad
-iV_Q'(i\chi)\in\mathbb{R}.
\label{eq:potential_reality_condition_axis}
\end{equation}
A sufficient condition is the anti-linear reality condition
\begin{equation}
V_Q(\varphi)^\ast=V_Q(-\varphi^\ast).
\label{eq:antilinear_reality_condition}
\end{equation}
This is weaker than imposing a full PT symmetry on the boundary theory, but it is enough to keep
the classical brane equations real on the ansatz \eqref{eq:imaginary_scalar_real_saddle}.

Indeed, substituting $\varphi=i\chi$ into the scalar equation gives
\begin{equation}
2\nabla_Q^2\chi=-iV_Q'(i\chi),
\label{eq:chi_EOM_real}
\end{equation}
which is real under \eqref{eq:potential_reality_condition_axis}.  Similarly, the Neumann condition
becomes
\begin{equation}
K_{ab}-K h_{ab}
+
\left(
V_Q(i\chi)-h^{cd}\partial_c\chi\partial_d\chi
\right)h_{ab}
+
2\partial_a\chi\partial_b\chi
=0,
\label{eq:Neumann_imaginary_scalar_real}
\end{equation}
which is again a real equation for a real embedding $X(z)$.

Thus the calculation below should be understood as a statement about this real classical saddle of
a complexified boundary theory.  It is not a statement about arbitrary non-Hermitian boundary RG
flows, nor does it imply that the full non-Hermitian spectrum is real.

%%%%%%%%%%%%%%%%%%%%%%%%%%%%%%%%%%%%%%%%%%%%%%%%%%%%%%%%%%%%%%%%%%%%%%%%%%%%%%%%%%%%%%%%%%%%%%
\subsection{Reversal of the holographic $g$-theorem}
\label{sec:ComplexBoundaryRGFlows:gtheorem}
%%%%%%%%%%%%%%%%%%%%%%%%%%%%%%%%%%%%%%%%%%%%%%%%%%%%%%%%%%%%%%%%%%%%%%%%%%%%%%%%%%%%%%%%%%%%%%

Let $k^a$ be a null vector tangent to $Q$.  Contracting
\eqref{eq:Neumann_complex_flow} with $k^ak^b$, the potential term drops out because
$h_{ab}k^ak^b=0$.  We obtain
\begin{equation}
\left(K_{ab}-Kh_{ab}\right)k^ak^b
=
2\left(k^a\partial_a\varphi\right)^2 .
\label{eq:null_contraction_complex_flow}
\end{equation}
For the embedding \eqref{eq:EOW_embedding_complex_flow}, one may choose
\begin{equation}
k^a\partial_a
=
\partial_t
+
\frac{1}{\sqrt{1+X'(z)^2}}\partial_z ,
\label{eq:null_vector_complex_flow}
\end{equation}
which gives
\begin{equation}
\left(K_{ab}-Kh_{ab}\right)k^ak^b
=
-\frac{X''(z)}
{z\left(1+X'(z)^2\right)^{3/2}},
\label{eq:geometric_null_contraction_complex_flow}
\end{equation}
with the orientation chosen so that a real scalar gives the usual holographic
$g$-theorem.

For a real scalar field, the right-hand side of
\eqref{eq:null_contraction_complex_flow} is non-negative, and hence
\begin{equation}
X''(z)\leq0.
\label{eq:real_scalar_g_theorem_concavity}
\end{equation}
This is the geometric positivity condition behind the holographic proof of the $g$-theorem.

For the imaginary saddle \eqref{eq:imaginary_scalar_real_saddle}, however,
\begin{equation}
\left(k^a\partial_a\varphi\right)^2
=
-
\left(k^a\partial_a\chi\right)^2 .
\label{eq:imaginary_scalar_negative_square}
\end{equation}
Equations \eqref{eq:null_contraction_complex_flow} and
\eqref{eq:geometric_null_contraction_complex_flow} then imply
\begin{equation}
X''(z)\geq0,
\label{eq:complex_scalar_g_theorem_convexity}
\end{equation}
with strict inequality whenever the imaginary scalar has a nonzero radial gradient.

The holographic boundary entropy function is
\begin{equation}
\log g(z)
=
\frac{1}{4G_N}
\operatorname{arcsinh}\frac{X(z)}{z}
=
\frac{c}{6}
\operatorname{arcsinh}\frac{X(z)}{z},
\qquad
c=\frac{3}{2G_N}.
\label{eq:holographic_g_function_real_saddle}
\end{equation}
At a fixed point $X(z)=z\sinh\rho_\ast$, this reduces to
\begin{equation}
\log g_\ast=\frac{\rho_\ast}{4G_N}=\frac{c}{6}\rho_\ast.
\label{eq:fixed_point_log_g_real_saddle}
\end{equation}
Differentiating \eqref{eq:holographic_g_function_real_saddle}, we find
\begin{equation}
\frac{d\log g}{d\log z}
=
\frac{1}{4G_N}
\frac{zX'(z)-X(z)}
{\sqrt{z^2+X(z)^2}},
\label{eq:g_derivative_real_saddle}
\end{equation}
and
\begin{equation}
\frac{d}{dz}
\left(zX'(z)-X(z)\right)
=
zX''(z).
\label{eq:zXp_minus_X_identity_real_saddle}
\end{equation}
If the flow starts from a UV fixed point,
\begin{equation}
X(z)=z\sinh\rho_{\rm UV}+o(z),
\qquad z\to0,
\label{eq:UV_fixed_point_real_saddle}
\end{equation}
then $zX'(z)-X(z)\to0$ in the UV.  Using
\eqref{eq:complex_scalar_g_theorem_convexity}, we obtain
\begin{equation}
zX'(z)-X(z)\geq0.
\label{eq:zXp_minus_X_positive_real_saddle}
\end{equation}
Therefore
\begin{equation}
\frac{d\log g}{d\log z}\geq0 .
\label{eq:g_increases_real_saddle}
\end{equation}
Thus the boundary entropy increases along the complex boundary RG flow:
\begin{equation}
\log g_{\rm IR}>\log g_{\rm UV}.
\label{eq:g_theorem_violation_real_saddle}
\end{equation}

This is a genuine reversal of the usual monotonicity statement at the level of the real classical
saddle: $X(z)$ is real and $\log g$ is real.  The reversal is not due to an ambiguity in ordering
complex numbers.  It follows from the loss of the positivity input in the null-energy step of the
standard proof.

%%%%%%%%%%%%%%%%%%%%%%%%%%%%%%%%%%%%%%%%%%%%%%%%%%%%%%%%%%%%%%%%%%%%%%%%%%%%%%%%%%%%%%%%%%%%%%
\subsection{Solution-generating map}
\label{sec:ComplexBoundaryRGFlows:solution_map}
%%%%%%%%%%%%%%%%%%%%%%%%%%%%%%%%%%%%%%%%%%%%%%%%%%%%%%%%%%%%%%%%%%%%%%%%%%%%%%%%%%%%%%%%%%%%%%

The real-saddle restriction above can be implemented constructively.  Let
\begin{equation}
\left(
X_R(z),\phi_R(z),V_R(\phi)
\right)
\label{eq:real_seed_solution_complex_flow}
\end{equation}
be any real solution of the brane-scalar system.  Define
\begin{equation}
X_C(z)=-X_R(z),
\qquad
\varphi_C(z)=i\phi_R(z),
\qquad
V_C(\varphi)=-V_R(-i\varphi).
\label{eq:complex_solution_map}
\end{equation}
The induced metric on $Q$ is unchanged by $X_R\to -X_R$, while the extrinsic-curvature
combination $K_{ab}-Kh_{ab}$ changes sign.  Meanwhile
\begin{equation}
h^{ab}\partial_a\varphi_C\partial_b\varphi_C
=
-
h^{ab}\partial_a\phi_R\partial_b\phi_R,
\qquad
V_C(\varphi_C)=-V_R(\phi_R).
\label{eq:scalar_terms_sign_flip_map}
\end{equation}
Thus the matter terms in \eqref{eq:Neumann_complex_flow} change sign in precisely the same way,
and the Neumann condition is mapped to itself.

The scalar equation is also mapped to itself.  Since
\begin{equation}
V_C'(\varphi)=iV_R'(-i\varphi),
\label{eq:potential_derivative_map}
\end{equation}
we have, on $\varphi_C=i\phi_R$,
\begin{equation}
2\nabla_Q^2\varphi_C
=
i\,2\nabla_Q^2\phi_R
=
iV_R'(\phi_R)
=
V_C'(\varphi_C).
\label{eq:scalar_EOM_map_check}
\end{equation}
Therefore \eqref{eq:complex_solution_map} maps ordinary real holographic boundary RG flows to
complex boundary RG saddles with real embeddings.  The real bulk geometry is not assumed by hand;
it is inherited from a real seed solution.

At fixed points, the map sends
\begin{equation}
\rho_R\longmapsto -\rho_R,
\label{eq:rho_sign_flip_map}
\end{equation}
and hence
\begin{equation}
\log g_C(z)=-\log g_R(z).
\label{eq:g_sign_flip_map}
\end{equation}
Since the ordinary unitary seed flow obeys
\begin{equation}
\log g_R^{\rm IR}<\log g_R^{\rm UV},
\label{eq:ordinary_g_theorem_seed}
\end{equation}
the complexified flow obeys
\begin{equation}
\log g_C^{\rm IR}>\log g_C^{\rm UV}.
\label{eq:complex_g_theorem_reversed_by_map}
\end{equation}
This provides a solution-generating proof of the failure of the boundary $g$-theorem in complex
BCFTs.

%%%%%%%%%%%%%%%%%%%%%%%%%%%%%%%%%%%%%%%%%%%%%%%%%%%%%%%%%%%%%%%%%%%%%%%%%%%%%%%%%%%%%%%%%%%%%%
\subsection{Defect entropy as a folded corollary}
%%%%%%%%%%%%%%%%%%%%%%%%%%%%%%%%%%%%%%%%%%%%%%%%%%%%%%%%%%%%%%%%%%%%%%%%%%%%%%%%%%%%%%%%%%%%%%
The same observation immediately gives a defect version.  By the folding trick, a conformal defect
$\ca{D}$ in a two-dimensional CFT is equivalent to a conformal boundary condition $B_{\ca{D}}$ in
the folded theory
\begin{equation}
\ca{C}_{\rm fold}
=
\ca{C}_L\otimes \overline{\ca{C}_R}.
\label{eq:folded_CFT_defect_entropy}
\end{equation}
The defect entropy is the Affleck--Ludwig entropy of this folded boundary:
\begin{equation}
\log g_{\ca{D}}
=
\log g_{B_{\ca{D}}}.
\label{eq:defect_entropy_as_boundary_entropy}
\end{equation}
Applying the complex solution map \eqref{eq:complex_solution_map} in the folded
AdS/BCFT description gives
\begin{equation}
\log g_{\ca{D}}^{\rm IR}>\log g_{\ca{D}}^{\rm UV}.
\label{eq:defect_entropy_violation}
\end{equation}
This is not a new independent mechanism; it is simply the folded version of
\eqref{eq:complex_g_theorem_reversed_by_map}.

%%%%%%%%%%%%%%%%%%%%%%%%%%%%%%%%%%%%%%%%%%%%%%%%%%%%%%%%%%%%%%%%%%%%%%%%%%%%%%%%%%%%%%%%%%%%%%
\subsection{Complex critical points and dS$_2$ branes}
\label{sec:ComplexBoundaryRGFlows:dSbrane}
%%%%%%%%%%%%%%%%%%%%%%%%%%%%%%%%%%%%%%%%%%%%%%%%%%%%%%%%%%%%%%%%%%%%%%%%%%%%%%%%%%%%%%%%%%%%%%

The fixed-point interpretation of $V_Q(\varphi_\ast)$ gives a precise sufficient
condition for realizing a de Sitter EOW brane from a complex boundary
deformation.  The point is not that an arbitrary complex RG trajectory should be
extrapolated beyond the usual AdS/BCFT regime.  Rather, whenever the complexified
brane-scalar theory has a real-saddle critical point with supercritical effective
tension, the full equations of motion themselves contain the de Sitter brane as
an exact solution.

Let us spell this out directly in the Lorentzian planar description.  At a fixed point,
\begin{equation}
\partial_a\varphi=0,
\qquad
V_Q'(\varphi_\ast)=0,
\end{equation}
the equation of motion for the scalar field is automatically satisfied, and the Neumann condition
\eqref{eq:Neumann_complex_flow} reduces to
\begin{equation}
K_{ab}-K h_{ab}+T_\ast h_{ab}=0,
\qquad
T_\ast=V_Q(\varphi_\ast).
\label{eq:constant_scalar_exact_tension}
\end{equation}
Thus the only input from the complex deformation at the fixed point is the real
number $T_\ast$.  We now look for a planar Lorentzian EOW brane of the form
\begin{equation}
Q_\lambda:\qquad t=\lambda z .
\label{eq:Lorentzian_dS_brane_embedding}
\end{equation}
For $|\lambda|>1$, the induced metric on this brane is Lorentzian,
\begin{equation}
ds_Q^2
=
\frac{L^2}{z^2}
\left(dx^2-(\lambda^2-1)dz^2\right).
\label{eq:dS_induced_metric_z}
\end{equation}
A direct evaluation of the extrinsic curvature shows that the fixed-point condition
\eqref{eq:constant_scalar_exact_tension} is solved provided
\begin{equation}
T_\ast L=-\frac{\lambda}{\sqrt{\lambda^2-1}},
\qquad |T_\ast|L>1 .
\label{eq:Lorentzian_supercritical_tension}
\end{equation}
Thus the ansatz \eqref{eq:Lorentzian_dS_brane_embedding} directly describes the
supercritical branch,
\begin{equation}
|T_\ast|L>1 .
\label{eq:supercritical_condition_exact}
\end{equation}
This is the spacelike-boundary branch of AdS/BCFT, known from analytic continuation
of Euclidean AdS/BCFT and appearing, for example, in the Lorentzian wedge construction
of Ref.~\cite{Akal2020}.  With
\begin{equation}
\eta=\sqrt{\lambda^2-1}\,z,
\end{equation}
the induced metric becomes
\begin{equation}
ds_Q^2
=
\frac{L_{\rm dS}^2}{\eta^2}
\left(-d\eta^2+dx^2\right),
\qquad
L_{\rm dS}
=
L\sqrt{\lambda^2-1}
=
\frac{L}{\sqrt{T_\ast^2L^2-1}} .
\label{eq:dS2_induced_metric}
\end{equation}
Therefore the implication
\begin{equation}
|T_\ast|L>1
\quad\Longrightarrow\quad
Q_\ast\simeq dS_2
\label{eq:sufficient_condition_dS_endpoint}
\end{equation}
is an exact statement about the classical saddle.

The sign of $T_\ast$ determines which supercritical branch is realized.  With the convention in \eqref{eq:Lorentzian_supercritical_tension},
\begin{equation}
T_\ast L<-1
\end{equation}
is the small-wedge branch naturally interpreted as a Lorentzian CFT with a
spacelike boundary, while
\begin{equation}
T_\ast L>1
\end{equation}
is the large-wedge branch whose holographic interpretation includes additional
degrees of freedom associated with gravity on dS$_2$ \cite{Akal2020}.

We now give an explicit brane potential satisfying the real-saddle condition and
producing such a supercritical fixed point.  Set $L=1$ and take
\begin{equation}
V_Q(\varphi)
=
T_0
-\frac12 M^2\varphi^2
-\frac14 g\varphi^4,
\qquad
T_0=\frac12,
\quad
M^2=\frac14,
\quad
g=\frac{1}{64}.
\label{eq:toy_potential_supercritical}
\end{equation}
Because the potential is even with real coefficients, it obeys the anti-linear
condition \eqref{eq:antilinear_reality_condition}.  On the imaginary axis,
\begin{equation}
V_Q(i\chi)=\frac12+\frac18\chi^2-\frac{1}{256}\chi^4,
\qquad
-iV_Q'(i\chi)=-\frac14\chi+\frac{1}{64}\chi^3,
\label{eq:toy_potential_reality_axis}
\end{equation}
so the scalar equation and Neumann condition are real on the ansatz
$\varphi=i\chi$.

This potential has imaginary critical points
\begin{equation}
\varphi_\ast=\pm i\frac{M}{\sqrt g}=\pm4i,
\qquad
V_Q'(\varphi_\ast)=0,
\end{equation}
and at either of them
\begin{equation}
T_\ast=V_Q(4i)=\frac32.
\label{eq:toy_supercritical_value}
\end{equation}
This is supercritical.  The corresponding exact EOW brane is
\begin{equation}
t=-\frac{3}{\sqrt5}\,z
\end{equation}
for the orientation with $T_\ast=3/2$, and its induced geometry is
\begin{equation}
ds_Q^2
=
\frac{4}{5\eta^2}(-d\eta^2+dx^2),
\qquad
L_{\rm dS}=\frac{2}{\sqrt5}.
\label{eq:toy_explicit_dS_brane}
\end{equation}
Equations \eqref{eq:toy_potential_supercritical}--\eqref{eq:toy_explicit_dS_brane}
are the promised realization: a complex relevant boundary deformation has a
real-saddle critical point at which the full AdS/BCFT brane equations are solved
by a de Sitter EOW brane.
\footnote{
This statement should be understood in the complexified saddle space.
For the toy potential, $V_Q(i\chi)$ indeed moves from the subcritical value $1/2$ to the supercritical value $3/2$, but this is not a single non-degenerate real Lorentzian EOW brane.
}

Note that the complex scalar does not provide a new Lorentzian construction of an isolated dS$_2$ brane; rather, it embeds the supercritical branch, which is absent from real Euclidean AdS/BCFT saddles with $|T|L<1$, as a complex critical point of the Euclidean boundary-condition space.

%%%%%%%%%%%%%%%%%%%%%%%%%%%%%%%%%%%%%%%%%%%%%%%%%%%%%%%%%%%%%%%%%%%%%%%%%%%%%%%%%%%%%%%%%%%%%%
%%%%%%%%%%%%%%%%%%%%%%%%%%%%%%%%%%%%%%%%%%%%%%%%%%%%%%%%%%%%%%%%%%%%%%%%%%%%%%%%%%%%%%%%%%%%%%
\section{Discussion}
\label{sec:Discussion}
%%%%%%%%%%%%%%%%%%%%%%%%%%%%%%%%%%%%%%%%%%%%%%%%%%%%%%%%%%%%%%%%%%%%%%%%%%%%%%%%%%%%%%%%%%%%%%
%%%%%%%%%%%%%%%%%%%%%%%%%%%%%%%%%%%%%%%%%%%%%%%%%%%%%%%%%%%%%%%%%%%%%%%%%%%%%%%%%%%%%%%%%%%%%%
We conclude by listing comments, open questions, and future directions.

\begin{itemize}

\item
It would be interesting to use complex conformal manifolds as analytic
benchmarks for the nonunitary conformal bootstrap; see, for example,
\cite{Huang2025}.  Unlike in the unitary bootstrap, positivity of OPE
coefficients is not available as a general input in
nonunitary CFTs.  It is therefore important to have solvable nonunitary
families whose conformal data are known independently.  The complex conformal
manifolds constructed in this work provide such examples: by analytically
continuing exactly marginal couplings, one obtains continuous families of
conformal data, often at fixed central charge, while scaling dimensions and
other universal quantities become genuinely complex.  These exact data can be
used to test bootstrap assumptions, truncation schemes, and numerical
strategies beyond the unitary regime.

\item
In this work, we have explicitly demonstrated that a unitary bulk CFT can support complex boundary and defect fixed points after analytic continuation of exactly marginal boundary/defect couplings.  This indicates that the boundary/defect phase diagram of an otherwise unitary CFT may contain many additional nonunitary fixed points that are not visible within the conventional real parameter space. For example, in the non-Hermitian Kondo problem, the interplay between dissipation and electron-electron interactions was argued to generate complex fixed points with spiral RG trajectories and walking behavior \cite{Han:2023ygh}, suggesting that complex fixed points can emerge robustly in impurity systems once the coupling space is enlarged. Also see recent exploration on complex defects in the $SU(2)_1$ Wess-Zumino-Witten CFT \cite{Kattel:2026cbp}.  Similarly, studies of Wilson lines in conformal gauge theories reveal nontrivial defect fixed-point structure, including fixed-point mergers, and screening-induced RG flows \cite{Aharony:2022ntz,Aharony:2023amq}. These examples strongly suggest that many unitary critical bulk systems may admit hidden complex boundary/defect fixed points, and that uncovering them could provide a unified language for nonunitary boundary criticality across condensed matter and gauge theory settings.

\item
In Sec.~\ref{sec:rational}, we have noted that, if one assumes rationality for a bulk CFT, then the central charge and conformal dimensions must necessarily be rational numbers. Here we briefly discuss whether an analogous statement also holds for CFTs with boundaries. The conclusion is that, if rationality is defined appropriately, even a complexified CFT can possess rational boundaries. In other words, although the bulk Hilbert space is, in general, an infinite direct sum, possibly continuous, of representations of a chiral algebra or of the Virasoro algebra, it is possible to construct a situation in which the open-string Hilbert space associated with the boundary decomposes into a finite direct sum.

A construction of this type, based on Liouville theory, was discovered long ago by Zamolodchikov and Zamolodchikov\cite{Zamolodchikov:2001ah}. In this construction, one imposes, at the two ends of the open string, boundary conditions corresponding to degenerate representations of the Virasoro algebra labeled by positive integers $(m,n)$. Then the partition function on the open-string Hilbert space decomposes into a finite sum of degenerate Virasoro characters as
\begin{equation}
    \mathcal{Z}_{(m,n)(m’,n’)}(q)=\sum_{k=0}^{\text{min}(m,m’)-1}\sum_{l=0}^{\text{min}(n,n’)-1}\chi_{m+m’-2k-1,n+n’-2l-1}(q).
\end{equation}
This construction also remains valid for complexified Liouville theory. It therefore shows that, even in a complexified CFT, the boundary sector can become rational in the above sense. For complexified Liouville theory, see \cite{Collier:2024kmo,Collier:2024mlg}.

This phenomenon admits the following mathematical interpretation. The category $\mathcal{C}$ describing the bulk theory is, in general, an infinite braided tensor category. Boundary conditions are then described by a module category $\mathcal{M}\simeq \mathrm{Mod}(\mathcal{C})$, and one chooses objects $(V_a,V_b,\ldots)$ of this category as boundary conditions. The open-string Hilbert space can equivalently be regarded as the space of boundary-condition-changing operators. Mathematically, this is the internal Hom object $\underline{\text{Hom}}_{\mathcal{M}}(V_a,V_b)$. If this object decomposes into finitely many simple objects, then one may say that the theory has a rational boundary.

\item
A potentially interesting connection is with the ``imaginary distance bound'' recently proposed for holographic CFTs, motivated by scalar-field wormholes \cite{Maldacena2026}. In our compact-boson example, the consistency domain (\ref{eq:bulkstrip}) precisely agrees with the proposed bound.
This agreement may provide a useful clue toward understanding a CFT-side derivation of the imaginary distance bound.
More broadly, it would be interesting to investigate the extent to which this holographically motivated bound continues to hold beyond holographic CFTs, and whether it reflects a more general constraint on conformal field theories.

\end{itemize}

\section*{Note added}
%{\textbf{Note added}} 
During the preparation of this work, we became aware of a related work \cite{Tang2026}, which will appear on arXiv on the same day. We thank the authors for coordination.

%%%%%%%%%%%%%%%%%%%%%%%%%%%%%%%%%%%%%%%%%%%%%%%%%%%%
\section*{Acknowledgments}
%%%%%%%%%%%%%%%%%%%%%%%%%%%%%%%%%%%%%%%%%%%%%%%%%%%%
We thank Nathan Benjamin, Brian McPeak, Slava Rychkov and Ning Su for discussions.
YK is supported by the INAMORI Frontier Program at Kyushu University and JSPS KAKENHI Grant Number 23K20046.
WH is supported by JST SPRING, Japan Grant Number JPMJSP2136.

\clearpage
\bibliographystyle{JHEP}
\bibliography{main}

\end{document}